\documentclass[a4paper,10pt,oneside]{article}

\usepackage{caption}
\usepackage{tgpagella}
\usepackage[T1]{fontenc}
\usepackage{latexsym}
\usepackage{amssymb}
\usepackage{amsmath}
\usepackage{float}
\usepackage{mathtools}
\usepackage[table]{xcolor}
\usepackage{mathrsfs}
\usepackage{multicol}
\usepackage{array}
\usepackage{longtable}
\usepackage{booktabs}
\usepackage{tabularx}
\usepackage{multirow}
\usepackage{enumitem}
\usepackage{booktabs}
\usepackage[most]{tcolorbox}
\usepackage{makecell}
\usepackage{colortbl}
\usepackage{hhline}
\usepackage{array,booktabs,arydshln,xcolor}
\usepackage{enumitem}

\usepackage{tikz}
\usetikzlibrary{arrows,decorations.markings}

\usepackage{theorem}

\usepackage{hyperref}

\usepackage{pgf}
\usepackage{xxcolor}

\usepackage{natbib}

\theoremheaderfont{\scshape}
\theorembodyfont{\rmfamily}

\bibpunct[, ]{(}{)}{,}{a}{}{;}

\def\qed{\relax
   \ifmmode
    ~\hfill\Box
   \else
    \unskip\nobreak ~\hfill$\square$%
   \fi \par}

\newcommand{\sep}{$\cdot$ }

\newcounter{namedlistcounter}  

\newcolumntype{M}[1]{>{\centering\arraybackslash}m{#1}}

\newcommand\tikzmark[2]{%
	\tikz[remember picture,overlay] 
	\node[inner sep=0pt,outer sep=0pt] (#1){#2};%
}

\newcommand\linka[2]{%
	\begin{tikzpicture}[remember picture, overlay, >=stealth, shorten >= 3pt,line width=1.5pt,blue]
		\draw [decoration={markings,mark=at position 1 with
			{\arrow[scale=1.75,blue,line width=0.5mm]{>}}},postaction={decorate}] (#1.east) to  (#2.west);
	\end{tikzpicture}%
}

\newcommand\linkb[2]{%
	\begin{tikzpicture}[remember picture, overlay, >=stealth, shorten >= 3pt,line width=1.5pt,blue]
		\draw [decoration={markings,mark=at position 1 with
			{\arrow[scale=1.75,blue,line width=0.5mm]{>}}},postaction={decorate}] (#1.east) to  (#2.west);
	\end{tikzpicture}%
}

\newcommand\linkc[2]{%
	\begin{tikzpicture}[remember picture, overlay, >=stealth, shorten >= 3pt,line width=1.5pt,blue]
		\draw [decoration={markings,mark=at position 1 with
			{\arrow[scale=1.75,blue,line width=0.5mm]{>}}},postaction={decorate}] (#1.west) to  (#2.east);
	\end{tikzpicture}%
}

\definecolor{Gray}{gray}{0.85}
\definecolor{LightGray}{gray}{0.90}
\definecolor{LightCyan}{rgb}{0.88,1,1}

\newcolumntype{a}{>{\columncolor{Gray}}c}
\newcolumntype{b}{>{\columncolor{white}}c}

\begin{document}
	\title{On the Nature of Discrete Space-Time \\ 
		\large Part 2:  Special Relativity in Discrete Space-Time}
    
	\author{DAVID CROUSE}
	\maketitle

\begin{abstract}
		In this paper--\textit{Part 2} of our series on discrete spacetime--we first provide a review of the previously published \textit{Part 1} that included the first important steps in the development of a new model of discrete spacetime (DST): the \textit{Isotropic Model} of DST.  The review starts with a description of the most important and oft-cited problems associated with DST models and how the \textit{Isotropic Model} of DST solves these problems.  We then summarize the derivations of the equations describing time dilation, length contraction and spatial distances in DST that were developed in \textit{Part 1}.  Following the review is the main part of the paper that includes the development of the DST versions of the Lorentz transformation equations and the derivation of the relativistic velocity addition rule, the construction of Minkowski spacetime in DST, and the resolution of Zeno's Stadium Paradox.  Important physical phenomena unique to the \textit{Isotropic Model} are discussed:  the fact that light experiences space and time in DST, and the uncertainty in the position of a particle upon the sudden termination of its travel.
	
	
\end{abstract}

\section{\label{sec:intro}Introduction}
What is space, what is time, motion, inertia...?  Which aspects of nature are fundamental and which properties are emergent at small or large size scales?  These are some of the questions involved in the 2500-year-old debate on the nature of space and time.  One particularly important question is whether space and time are continuous or whether they come in discrete steps.  Most people cite the writings of Parmenides (c. 500 BCE) and the paradoxes of his mentee Zeno of Elea as the naissance of this debate (see \cite{Salmon2001,Owen1957,Grunbaum1952,Dowden2020} and \cite{Hugget2018}).  And while interest in discrete spacetime (DST) may have lessened for the next couple of millennia (with a few notable exceptions (e.g., \cite{Maimonides1190})), there has been a resurgence of interest in DST. This is due primarily to the large community of contemporary scientists who are developing physical theories that aim to describe natural phenomena at extremely small scales, e.g., quantum mechanics with \cite{Hagar2014, March1951,Gudder2017} and \cite{Ruark1931}; causal set theory with \cite{Henson2008, Rideout2009} and \cite{Reid1999}; and quantum loop gravity with \cite{Rovelli2003}. Much fewer in number are people who have recently been drawn to the topic to develop a philosophically and mathematically consistent model of discrete spacetime and motion, namely \cite{Salmon2001,Crouse2019,VanBendegem1995,Forrest1995} and \cite{Harrison1996} to name a few. I refer the reader to \textit{Part 1} of this work (\citet{Crouse2019}) and Amit Hagar's work (\cite{Hagar2014}) for further discussion on the history of DST\textemdash the purpose of this paper is to finish the development of special relativity in DST and present an analysis and solution to Zeno's Stadium Paradox. 

This paper is organized as follows.  To provide context for the rest of this work, the problem-plagued conventional model of DST is first described in the \textit{Introduction}.  Importantly, descriptions are given of the major problems inherent to most DST models that should be addressed in any quality work on the topic.  Section \ref{sec:Review} contains a review of the \textit{Isotropic Model} that was developed in \textit{Part 1} (\citet{Crouse2019}).  The new work of this paper starts in Section \ref{sec:Space_Transformation}, where the Lorentz transformation equations for the spatial coordinates in DST are derived.  Section \ref{sec:Clock_Syncrohnization} starts with a review of the well-known phenomenon of time asynchronicity in continuous spacetime (CST) and is followed by a description of how one properly calculates time asynchronicity in DST.  Following this is the derivation of the Lorentz transformation equations for time coordinates in DST.  The DST versions of the Lorentz transformation equations are given in Section \ref{sec:Lorentz_DST} and the relativistic velocity addition rule in Section \ref{sec:Velocity_Addition}.  Minkowski DST spacetime diagrams are described and constructed in Section \ref{sec:Minkowski}.  We then discuss and resolve Zeno's Stadium Paradox in Section \ref{sec:Stadium}.   Upon completing all of the derivations and descriptions of the foundational aspects of DST in the prior sections, we transition in Section \ref{sec:SRTopics} to a discussion of some of the most interesting and unique characteristics of, and phenomena exhibited within DST.  The first consequence is the ability for particles with nonzero mass to travel at a speed that is indistinguishable from that of light over some period of time if the particles have sufficient energy.  The second consequence is the unavoidable uncertainty that arises in DST of a particle's position upon the sudden termination of its travel.   Section \ref{sec:Conclusion} concludes the paper with a summary of its main results and a discussion of an easily observable phenomena that can \textit{only occur if spacetime is discrete}, namely light's ability to interact and undergo change.

\subsection{\label{sec:conventional}Conventional Model of Discrete Spacetime}

\begin{figure}
	\centering\includegraphics[width=6cm]{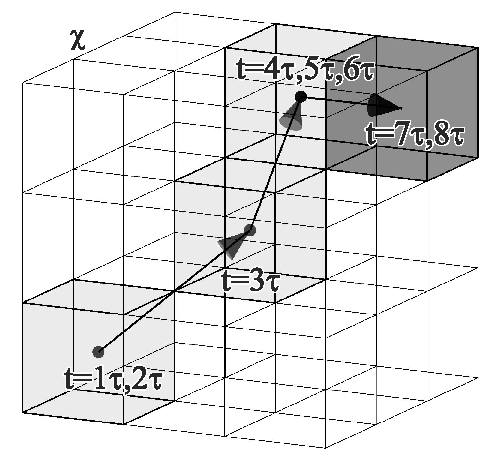}
	\caption{The conventional (\textit{but flawed}) model of DST as a lattice.  In these models, space is split up into a grid, or lattice, with cubic unit cells having a volume of $\chi^3$.  Time occurs as snapshots, each of duration $\tau$.  Motion in DST involves a particle (darker gray) either transitioning to an immediately neighboring cell upon a snapshot of time, or remaining at its location. An example path (light and dark gray cubes) of a particle traveling at an \textit{average velocity} of $c/2$ is shown.  The times shown indicate where the particle is at the end of a tick; for example, the particle is at the lower left cell for the continuous-time ranges of $t=0 \rightarrow 1\tau$ and $t=1\tau \rightarrow 2\tau$, then transitioning to the middle cell where it remains for only one chronon in time, i.e., $t=2\tau \rightarrow 3\tau$, and so on. In this paper we present a superior model---the \textit{Isotropic Model of DST}---that is not plagued by the problems associated with these lattice models.}
	\label{fig:Fig1}
\end{figure}

Most conventional models of discrete space have continuous space being segmentized (atomized, discretized, granularized...) in a grid-like way with exceptionally small cubic building-blocks called hodons (Fig. \ref{fig:Fig1}).  These building blocks have a volume of $\chi^3$, with $\chi$ being nature's ostensibly smallest spatial length, typically taken to be equal to the Planck length $\chi=l_p=1.62 \times 10^{-35}$ m.\footnote{Justification for a value for $\chi$ as $2l_p$ is given in \cite{Crouse2019}.}  Alternatively, these \textit{Planck volumes} can be tetrahedrons (\cite{Aybar2017}), Penrose tiles, or unit cells of other shapes. In such a system, an elementary particle at a particular instant in time is located within \textit{and throughout} any one cube.  For discrete time, the conventional models have time evolving not as a continuous stream but as a series of discrete snapshots.  The extent of continuous time within each snapshot is $\tau=c / \chi$, where $\tau$ (called a chronon) is nature's ostensibly smallest temporal duration, typically taken to be equal to the Planck time $\tau_p=5.39 \times 10^{-44}$ s.

Particle motion in the conventional model occurs from one snapshot to the next, with a particle either staying at a particular unit cell or moving a distance $\chi$ to an adjacent cell.\footnote{Of course, one must address the profound issues that this statement raises\textemdash issues that have been debated for over 2500 years via Zeno's paradoxes. Zeno's Arrow Paradox focuses on the topic of when movement occurs: during a snapshot or between snapshots?...both scenarios are unsatisfying in DST models. Another work by the author discusses all of Zeno's paradoxes that are pertinent to the DST debate, in this paper we only discuss in detail the Stadium Paradox.}  In the cubic lattice model shown in Fig. \ref{fig:Fig1}, an adjacent cell is any of the following types:  one connected to the original cell across a face of the cell, one connected along an edge of the cell, and one connected along the diagonal.  Examples of all three of these types of movements are shown in Fig. \ref{fig:Fig1}.\footnote{The reader will probably identify one key failing of this model at this point in its description, namely, these distances appear not to be the same: $\chi$ for transitions across a face, $\sqrt{2}\chi$ for transitions across an edge, and $\sqrt{3}\chi$ for transitions along a diagonal.  However, it was Herman Weyl in 1949 (\cite{Weyl1949}) who stated that all of these distances \textit{must be the same}, namely, the fundamental length $\chi$. And if this is true, Weyl argued that it necessarily follows that the diagonal must be equal to the sides of an equilateral right triangle of \textit{any size}. This statement has proven to be quite unsettling to philosophers and scientists since first penned by this philosophical and mathematical giant and is the basis of Weyl's \textit{tile-argument} against DST. We discuss the tile-argument and four other major problems typically associated with DST later in this section of the paper.}  Of course, all plausible DST models have the ratio of $\chi$ to $\tau$ being equal to $c$ (i.e., $\chi/\tau=c$) with $c$ being the speed of light. Hence, motion at the finest temporal resolution (i.e., one tick of duration $\tau$) involves a particle either at rest $v=0$ (when it \textit{has not} translated to a neighboring cell of any type) or traveling at the speed of light $v=\chi/\tau=c$ (when it \textit{has} translated to a neighboring cell).  Particles can have \textit{average} velocity values ($v_{avg}$) different than $v=0$ or $v=c$ over longer durations because $v_{avg}$ is the sum of the spatial translations divided by the total temporal duration ($T=N\tau$), i.e., $v_{avg}=(1/T)\sum_{i=0}^{N} d_i$, where $d_i$ is either $0$ or $\chi$, and $N$ is an integer. For example, if one states that a particle's velocity is $0.5c$ (as in Fig. \ref{fig:Fig1}), it really means that the average velocity $v_{avg}$ is $0.5c$ over time scales large relative to $\tau$; on the finest temporal resolution this means that the \textit{probability} that the particle undergoes a translation instead of remaining at its current lattice point is $0.5$. Before diving deeper into the mathematical and philosophical work of this paper let us draw an awe-inspiring mental picture of the tremendous smallness of $\chi$ and $\tau$.

Consider the following...\textit{the universe is to us what the hydrogen atom is to a Planck length}.  When comparing the relative lengths involved (\textit{i.e.,} universe to human versus hydrogen atom to Planck length) this statement is approximately true.  Take a moment to imagine each hydrogen atom being a \textit{Planck universe} inhabited by Planck-size citizens. A natural question to ask is: How well can we humans probe the inner workings of the Plancksonians' universe?  To answer this question, we consider humanity's current ability to probe physical phenomena at small size scales.  Our most advanced particle accelerators are able to probe properties of quarks and other subatomic particles at a size-scale of $10^{-19}$ m.  Unfortunately, this size-scale is still a factor of $10^{16}$ greater than the Planck length.  From this analysis we see that we are only able to probe features of this fictitious Planck universe that---scaling up to our human-size perspective---are equivalent in size to the distance from Earth to Proxima Centauri!  This realization should not lead one to despair.  I myself find it heartening to realize that the age of scientific discovery we are experiencing is far from over...why should we not expect that the panoply of physical phenomena observed and explained by us in the fields of chemistry, biology, physics, and cosmology for size scales greater than the hydrogen atom be matched by a similarly rich set of heretofore unimagined and awe-inspiring phenomena observed by our imagined Plancksonians...and by us humans in the future as our technology improves?! However, the entire concept of DST is fraught with profound and perhaps ruinous philosophical, physical, and mathematical problems.  In this work we believe that we provide solutions to most if not all of these problems---we start with a description of the problems in the next section.

\subsection{Plausibility Criteria for any DST Model}
It is well known that \textit{lattice-based} models of DST are highly problematic and, unfortunately, cause most physicists, philosophers and others to dismiss the entire DST concept outright. The numerous problems associated with DST raised by countless thinkers over the ages are discussed in \textit{Part 1} of this work (\cite{Crouse2019}) (and also by \cite{Hagar2014}) and can be reduced in number to the five important ones listed below. 

\begin{enumerate}
	\item\label{Problem1} \underline{Lorentz Contraction}:  According to special relativity, is not the ostensibly smallest possible unit of length (i.e., the hodon $\chi$) in one inertial reference frame Lorentz contracted to yet smaller lengths in moving reference frames, and similarly for the chronon $\tau$?
	
	\item\label{Problem2} \underline{Anisotopy}:  Is it unavoidable that discrete space will introduce preferred directions in space?  In other words, will not the motion of particles be dependent on the direction of travel, even in matter-free space?
	
	\item\label{Problem3} \underline{Calculation of Distances in DST}: This problem is best known via the famous Weyl-tile problem.  In 1949, Hermann Weyl claimed that if space is discrete, the length of the side of \textit{of any size} square must be equal to its diagonal.  This violation of the Pythagorean theorem is not observed, hence space must not be discrete (\cite{Weyl1949,Crouse2019}).
	
	\item\label{Problem4} \underline{Nonconservation of Energy and Momentum}: Motion of a particle in DST occurs through discrete jumps from one grid point to the next---something thought to be unphysical (\cite{VanBendegem2017a}).  Each single spatial jump (of length $\chi$) occurs over one fundamental duration of time (i.e. $\tau$). For particle motion with $v_{avg}<c$, any jump may be followed by a duration of some integer multiple of $\tau$ during which no jump occurs, and then followed by the next jump. Over the duration of each jump, the particle is presumably traveling at the speed of light $c=\chi / \tau$, with the particle having a nonzero kinetic energy; during nonjump durations the particle has zero kinetic energy. It seems that the law of conservation of energy and momentum is repeatedly violated in such a model!
	
	\item\label{Problem5} \underline{Zeno's Stadium Paradox}:  Amazingly, this 2500-year-old paradox still plays an important role in the debate of continuous versus discrete spacetime.  As described in Section \ref{sec:Stadium}, the contemporary version of the paradox, describe nicely by Adolf Grünbaum in \citep[248]{Salmon2001}, involves relative motions of three reference frames in DST that seemingly leads to the contradiction: ``half the time equals the whole time'' (\cite{VanBendegem2017b}).  This means that if one assumes a ``quantum of time'' $\tau$, the Stadium Paradox forces you to acknowledge the reality of $\tau/2$, and upon subsequent $n$ applications of the procedure described in the paradox, the reality of $\tau/2^n$ must be acknowledged (where $n$ is any natural number). 
\end{enumerate}
In order for any DST model to garner any degree of legitimacy, it must address all five of these problems.\footnote{Common quagmires encountered in the debate should be avoided. For example, questions such as ``what is the shape of space?'' and ``is not causality violated if a force applied to one side of a hodon is simultaneously experienced on the other side?''.  These questions indicate a failure to understand the basic nature and concept of DST, or a failing of the inquisitor to divorce themselves of the perspective of continuous spacetime.}  In \textit{Part 1} of our work (\cite{Crouse2019}), we discussed four models that seek to resolve some or all of these problems.  Three of these are lattice models proposed by Hermann Weyl (\cite{Weyl1949}), Jean Paul Van Bendegem (\cite{VanBendegem2017a}), and Peter Forrest (\cite{Forrest1995}) whose purposes were to address two issues:  calculating distances in DST and addressing the anisotropy issue in DST.   Their purposes were not to construct models that address all the problems listed above.  Hermann Weyl's model introduced the troublesome issue of the calculation of distances in DST, while Van Bendegem's (\cite{VanBendegem1995}) and Forrest's (\cite{Forrest1995}) models sought to resolve the Weyl-tile problem and minimize, but not eliminate the anisotropy problem. The non-lattice-based \textit{Isotropic Model} of DST developed by Crouse is superior to any lattice model, addressing and resolving the first three problems listed above; it addresses \textit{Problem 4} by arguing that inertia and mass (therefore kinetic energy and momentum) are not inherent properties of a particle, but emerge via numerous interactions with other particles over durations in time much larger than $\tau$, thereby obviating this problem (\cite{Crouse2019}).\footnote{The work of Ludwik Silberstein (\cite{Silberstein1937}) in 1937 should be mentioned since it has a couple of similarities to the \textit{Isotropic Model}.  His model includes an ordered array of points in three chosen directions, the fact that particle motion occurs point by point without skipping any intermediary point, and at least an attempt to marry DST with special relativity.  However, he did not use the correct distance formula, nor properly derive the Lorentz transformation equations, $\dots$.  One very compelling aspect of his work however, is his discussion of Taylor's theorem in DST.} As for the final problem, namely the Stadium Paradox, we resolve it in Section \ref{sec:Stadium} of this work.

\section{\label{sec:Review}Review of Part 1:  The Isotropic Model of DST}

The key aspect of the \textit{Isotropic Model of DST} developed by Crouse in \textit{Part 1} is correctly identifying what is spatially discretized: \textit{translations in any direction}.  Importantly, absent in the model is the imposition of a grid with fixed coordinates and directions.  This is in contrast to lattice models in which distances along some arbitrarily chosen $x$, $y$ and $z$ directions of a grid are discretized.  Time discretization in the \textit{Isotropic Model} is identical to that described above for lattice models.  A preliminary step was done first in \textit{Part 1}, namely developing a simple heuristic argument as to why space and time must be discretized, and then deriving the extent(duration) of the hodon(chronon). We then developed the model, first developing a formula for the calculation of distances in DST.  We then showed how to assign continuous spacetime $x$, $y$, and $z$ coordinates to points in the model.  Finally, we derived the DST versions of the equations describing Lorentz contraction and time dilation.  This last result yielded a resolution to the common but incorrect belief that DST, with it's immutable atoms of space and time, is incompatible with special relativity's Lorentz contraction and time dilation.\footnote{Carlo Rovelli also demonstrated a resolution of this DST/special relativity incompatibility using quantum loop gravity concepts (\cite{Rovelli2003}).} Specifically, we showed that the hodon and chronon are not contracted or dilated in any reference frame, regardless of its velocity relative to another reference frame.  Below, I summarize the important aspects of the model that are developed and discussed in detail in \textit{Part 1}.  

\subsection{The Extent of the Hodon and the Duration of the Chronon}

In \textit{Part 1}, we described three different approaches to calculate the extent and duration of the hodon and chronon respectively. The approaches used methods from causal set theory (CST) (\cite{Reid1999}), quantum loop gravity (QLG) (\citet{Rovelli2003}), and a simple heuristic approach based on minimal measurable spatial lengths and temporal durations (\cite{Crouse2019}).  All of these approaches yield similar values:  CST and QLG yield values of $\chi$ and $\tau$ being equal to $l_p$ and $\tau_p$ respectively, whereas the heuristic approach yielding $\chi=2l_p$ and $\tau=2\tau_p$.

\subsection{Coordinates and Distances in Discrete Spacetime}
With values for $\chi$ and $\tau$ in hand, we then developed in \textit{Part 1} expressions for the associations of intervals of positions \textit{along a straight path} in continuous space $\tilde{x}_m$ to their DST counterparts, and intervals of instances in continuous time $\tilde{t}_n$ to their DST counterparts (\cite{Crouse2019}):

\begin{subequations}
	\begin{align}
	\tilde{x}_m &= \left ( \left(m-\frac{1}{2} \right ) \chi, \left( m+\frac{1}{2} \right ) \chi \right ]\label{space_mapping} \\
	\tilde{t}_n &= \left [ n \tau, \left(n+1 \right) \tau \right )\label{time_mapping}
	\end{align}
\end{subequations}

\noindent with $m$ being any positive or negative integer and $n$ being an integer greater than or equal to zero. Thus, the continuous spacetime quantities $\tilde{x}_m$ and $\tilde{t}_n$ are always intervals, whereas their DST counterparts are single numbers.  The intervals defined by Eq. \eqref{space_mapping} are along \textit{any straight path}, not necessarily along any particular axis.

Following this, we needed to develop a method that assigns $x$, $y$ and $z$ coordinate values to positions within discrete space (\cite{Crouse2019}).  It was only at this point in the development of the model that we did what was anathema to us and what led many past DST models astray:  we drew axes.  However, we did so with full knowledge that these axes are \textit{arbitrary} and only a tool for us to assign coordinates to points in space and, importantly, they \textit{do not imbue the model with anisotropy}.  Why do we do this at all?  Because the assignment of coordinate values is necessary such that the model can be used as a foundation for mathematical and physical theories (e.g., general relativity and quantum field theory (\cite{Gudder2017})). The procedure is as follows.  First, choose and draw the desired set of axes (Fig. \ref{fig:Fig2}).  Following this, we state that Point $C$ is arrived at (starting from the origin Point $A$) by translating $m$ spatial atoms in the defined $x$ direction followed by a $n'$ spatial atom translation in the defined $y$ direction.  Thus, we assign the coordinates $(x_m=m\chi,y_{n'}=n'\chi)$ to this point (Point $C$). Following this procedure, one can associate all positions in CST to positions in DST in the correct way.

\subsection{The Discrete Spacetime Distance Formula}
\begin{figure}[H]
	\centering\includegraphics[width=7.5cm]{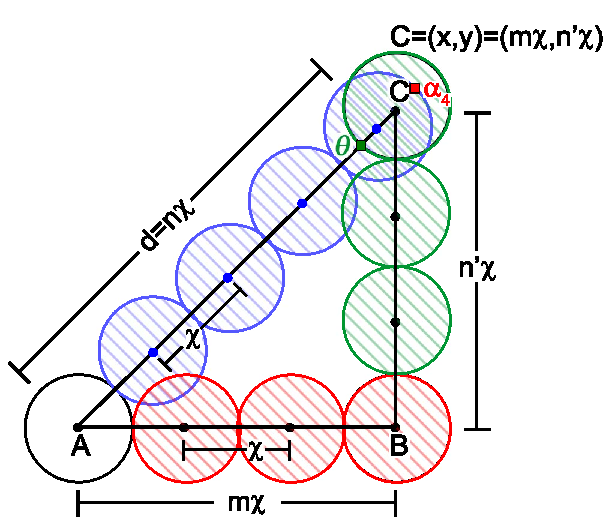}
	\caption{For any triangle, the distance formula Eq. \eqref{leopold2D} is derived by determining how many translations are required along $\overline{AC}$ such that the leading edge of the translating point (denoted by $\alpha_n$) along the hypotenuse overtakes the trailing edge (denoted by $\theta$) of the sphere that defines point $C$ (\cite{Crouse2019}).}
	\label{fig:Fig2}
\end{figure}

We were then in a position to address the old question of determining general distances in DST.  To do so, we looked at sequential \textit{and straight} translations from $A$ to $C$ along the diagonal connecting $A$ and $C$ and proposed the condition at which the translating point (in blue in Fig. \ref{fig:Fig2}) reaches $C$. The condition is: when any part of the translating point is at least partially within the point that defines $C$, then the translating point has reached $C$.  The distance (relative to $\chi$) from the origin to $C$ is easily calculated as:
\begin{equation}\label{leopold2D}
n = \left \lfloor \sqrt{m^2+n'^2} \right \rfloor 
\end{equation}
\noindent Again note that the axes were arbitrarily chosen---it is the translations in any direction that are discretized...this property of the model ensures that space \textit{remains isotropic} in this model of DST.  Also note that the floor operation is performed on the right side of Eq. \eqref{leopold2D}, doing so yields interesting but necessary violations of the triangle inequality theorem (necessary because of the operation's impact on relativistic length contraction) that are discussed in \textit{Part 1} of this work.\footnote{We will see in Section \ref{sec:TD_LC_DST} that the smallest light-clock requires the hypotenuse to be equal to one hodon when the base and height are both equal to one hodon.  Therefore the hypotenuse needs to skew smaller than the value given by the Pythagorean theorem rather than larger than it as the size of the triangle decreases to Planck-scale, hence the use of the floor operator in Eq. \eqref{leopold2D} rather than the ceiling operator.}$^,$\footnote{The only violations of the triangle inequality are when the legs of the triangle are colinear.  Consider points $A=(0,0)$, $B=(1,1)$, and $C=(3,3)$.  Using Eq. \eqref{leopold2D} to calculate distances, we have $\overline{AB}=1$, $\overline{BC}=2$, and $\overline{AC}=4$.  Thus, we do not have $\overline{AB}+\overline{BC}>\overline{AC}$.  This aspect of DST is essential for the model to adhere to the laws of special relativity. }  
\subsection{Time Dilation and Length Contraction in DST}\label{sec:TD_LC_DST}
\begin{figure}[H]
	\centering\includegraphics[width=0.75\textwidth]{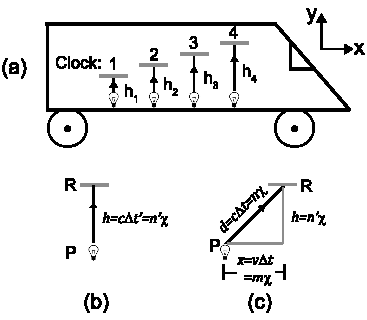}
	\caption{From \cite{Crouse2019} \textbf{(a):} An array of light clocks on a train traveling at a velocity $v$.  The clocks have values of $h$ as integer multiples of $\chi$.  \textbf{(b):} One of the clocks from the perspective of an observer in RF2.  \textbf{(c):} One of the clocks from the perspective of an observer in RF1.}.  
	\label{fig:Fig3}
\end{figure}

Now armed with the formula (Eq. \eqref{leopold2D}) for the distance of the hypotenuse of a triangle in DST, we were then able to perform the standard calculation to determine time dilation that is found in any introductory textbook on special relativity (\textit{e.g.,} \cite{Helliwell2010}).  Instead of just one clock, though, we used an array of clocks with different distances between their emitters and receivers---distances being integer multiples of $\chi$ (see Fig. \ref{fig:Fig3}).  This is necessary because the use of Eq. \eqref{leopold2D} instead of the standard Pythagorean theorem imparts a time dependence upon the Lorentz factor $\gamma=n/n'$ in addition to its velocity dependence.  This is in distinct contrast to the conventional form of the Lorentz factor in CST, namely $\gamma_{CST}=1/\sqrt{1-v^2/c^2}$, which depends only on the relative velocity of the moving reference frame. This aspect of $\gamma$ in DST, namely, its time dependence is of primary importance for the success of the model.

One can use two methods to calculate the quartets $(n,m,n',m')$:  the first method allows for the derivation of a simple formula for $\gamma$ in DST but have the hodon-in-extent translations occur at specific times, the second method is more complex to use but allows for hodon-in-extent translations to occur more realistically: randomly or probabilistically.

In the first method, we assume that the light-clock repeatedly moves at the end of certain time periods according the equation:

\begin{equation}
m = \bigg \lfloor \frac{v_{avg}}{c}n \bigg \rfloor
\end{equation}

\noindent For example, if $v_{avg}=0.25c$, then the light-clock will move one hodon on the last tick of every four-tick duration in RF1.  The sacrifice of the more realistic random movement is offset by the benefit of an easily-derived equation for $\gamma^{-1}$ acting upon any of member of the quartet $j=n,m,n',m'$:

\begin{equation}
	\gamma^{-1}_j j = \Biggl \lceil   \sqrt{j^2 - \Big\lfloor \frac{v_{avg}}{c}j \Bigr\rfloor^2 +2j+1} -1  \Biggr\rceil 
\end{equation}

\begin{figure}[H]
	\centering\includegraphics[width=11cm]{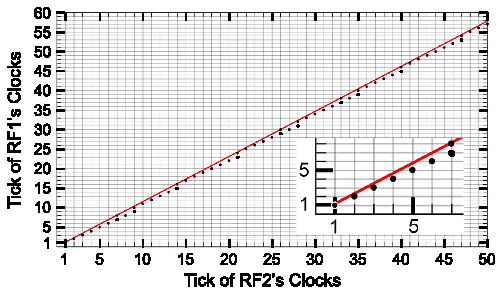}
	\caption{From \cite{Crouse2019}.  The correspondence between the ticks of the clocks in RF1 to the ticks of the clocks in RF2. The RFs have a relative average velocity of $v_{avg}=0.5c$.  The red line shows $\gamma_{CST}=1.15$.  \textbf{Inset}: No time dilation occurs for the first seven ticks (each of duration $\tau$), but then the clocks on the train start trailing the clocks at the station.}
	\label{fig:Fig4}
\end{figure}

The second method has you iteratively use Eq.\eqref{leopold2D} for each tick $n$ and having the freedom to set the movement of the light-clock to occur at regularly assigned ticks, randomly, bunched together, or anything else.  The procedure is described below, with results shown in Table 1 and Fig. \ref{fig:Fig4} for an example with $v_{avg}=v=0.5c$ and with the movement happening at the second tick of every 2-tick sequence.\footnote{See \textit{Part 1} for greater detail.  Matlab programs to implement this process and all other processes and calculations described in this work are available by contacting the author.}  Note that $n\tau$ and $m\chi$ are the time and length in the at-rest frame RF1 with $n$ and $m$ being integers; $n'\tau$ and $m'\chi$ are the time and length in the moving frame RF2 with $n'$ and $m'$ being integers.

\begin{enumerate}
	\item \underline{Given $v$ and $n'$, find $n$ and $m$}:  In this first step, one scans through $n'$ values, namely $n'=1,2,3,...$.  For each value of $n'$, one creates an array of $m$ values $m=0\rightarrow N$ (where $N$ is large).  One inserts the single value of $n'$ and the array or $m$ values into Eq.\eqref{leopold2D} and obtains a list of candidate $n$ values.  One then chooses the $(n,m)$ set with the desired distance traveled by the light-clock $m$ for that particular time $n$.  
	\item \underline{Use $n$, $m$, and $n'$ to find $m'$}:  One then uses the fact that the magnitude of the velocity $v$ is the same when viewed from either reference frame, namely, $|v|=m/n=m'/n'$.  The procedure is as follows for the example with $(n,n',m,m')=(30,27,15,13)$ in Table \ref{table:Table1}:  for a particular $n'$ value (orange in Table 1), find the $n$ that matches this value (shaded in purple) in an earlier row of the table; find the $m$ value in this row (shaded in pink).  This $m$ value is the $m'$ value (shaded in yellow) for the original row of data.  
	\item \underline{Use $n$ and $n'$ to find $\gamma$}:  To find the value of $\gamma$ for each value of $n'$ used in Step 1, one simply calculates the ratio $n/n'$; this ratio is the value of $\gamma=n/n'$ for that particular time (\textit{i.e.,} $n$ and $n'$).
\end{enumerate}

The data (i.e., $n$, $n'$, $v$, $\gamma$, $m$, and $m'$) for the first 33 ticks and 50 ticks of the RF1's clock are shown in Table 1 and Fig. \ref{fig:Fig4} respectively.

An important case that will be used later in this work when studying the Stadium Paradox is when the moving reference frame (RF2) is traveling at $v=c$ relative to the stationary reference frame (RF1).  In CST, the Lorentz factor is infinity, but in DST the Lorentz factor is finite and can be obtained by setting $m=n$ in Eq. \eqref{leopold2D}, resulting in the equation:
\begin{subequations}
	\begin{align}
	n &= \left \lfloor \frac{n'^2-1}{2} +1 \right \rfloor \xrightarrow{n' \gg 1}  \frac{n'^2}{2}  \label{light_speed} \\
	n'&= \left \lceil \sqrt{2n+1}-1 \right \rceil \xrightarrow{n \gg 1} \sqrt{2n}
	\end{align}
\end{subequations}
\noindent The Lorentz factor for this case (i.e., $v=c$) is:
\begin{equation}\label{light_speed_gamma}
\gamma =\frac{1}{n'} \left \lfloor \frac{n'^2-1}{2} +1 \right \rfloor = \frac{n}{\left \lceil \sqrt{2n+1}-1 \right \rceil}\xrightarrow{n,n' \gg 1} \frac{n'}{2}=\sqrt{\frac{n}{2}}
\end{equation}
\noindent  We see that $\gamma$ diverges to infinity as $n$ and $n'$ become large---in agreement with that predicted by the laws of special relativity in CST.  In DST, however, for any finite time $t=n\tau$ in the rest frame, the photon will experience an amount of time $t_{photon}=n'\tau$ given by:
\begin{equation} \label{photon_time}
t_{photon}= \gamma^{-1}_n t = n'\tau \xrightarrow{n' \gg 1} \sqrt{2n}\tau=\sqrt{2\tau t}
\end{equation}
\noindent We therefore conclude that photons share the flow of time with us, albeit to a much reduced extent (see Table \ref{table:Table2}).  This particular feature of special relativity in DST is of critical importance because it is in direct conflict with the CST version of special relativity and makes a strong argument for DST as being the true nature of space and time. The argument goes like this:  
\begin{quote}
\textit{Since photons of course experience change, as evidenced by their interactions with other systems (\textit{e.g.,} our eyes, lightbulbs, cameras, other photons via interference...) or as evidenced by change that they experience in isolation (e.g., particle-antiparticle creation), they must experience time.  Special relativity in CST seemingly predicts that photons do not experience any extent of our rest-frame time, whereas photons in the DST version of special relativity do experience it (see Eq. \eqref{photon_time}).  Therefore DST is a superior model for spacetime.}
\end{quote}
We are now in a position to continue with the new material in the development of special relativity in discrete space-time, namely determining clock asynchronicity, deriving the Lorentz transformation equations, deriving the velocity addition rule, construction of Minkowski spacetime, and solving Zeno's Stadium Paradox.

\begin{table}[H]
\centering
	\caption{\textit{Spreadsheet Mechanics}:  Clock tick associations, length associations, and Lorentz factors for $v_{avg}=0.5c$.}\label{table:Table1}
\begin{tabular}{|c|c|c|c|c|c|}
	\hline
	Hypotenuse ($n$) & Height ($n'$) & Base & Contracted & $\gamma_n$ & $\gamma_m$\\
	RF1's clock tick & RF2's clock tick & ($m$) & length (m') & $(n/n')$ & $(m/m')$\\
	\hline
	1 & 1 & 0 & 0 & 1 & - \\
	\hline
	2 & 2 & 1 & 1 & 1 & 1 \\
	\hline
	3 & 3 & 1 & 1 & 1 & 1 \\
	\hline
	4 & 4 & 2 & 2 & 1 & 1 \\
	\hline
	5 & 5 & 2 & 2 & 1 & 1 \\
	\hline
	6 & 6 & 3 & 3 & 1 & 1 \\
	\hline
	7 & 7 & 3 & 3 & 1 & 1 \\
	\hline
	8 & 7 & 4 & 3 & 1.14 & 1.33 \\
	\hline
	8 & 8 & 4 & 4 & 1 & 1 \\
	\hline
	9 & 9 & 4 & 4 & 1 & 1 \\
	\hline
	10 & 9 & 5 & 4 & 1.11 & 1.25 \\
	\hline
	11 & 10 & 5 & 5 & 1.10 & 1  \\
	\hline
	12 & 11 & 6 & 5 & 1.09  & 1.20 \\
	\hline
	13 & 12 & 6 & 6 & 1.08  & 1 \\
	\hline
	14 & 13 & 7 & 6 & 1.08  & 1.17 \\
	\hline
	15 & 14 & 7 & 7 & 1.07 & 1 \\
	\hline
	16 & 14 & 8 & 7 & 1.14 & 1.14 \\
	\hline
	17 & 15 & 8 & 7 & 1.13 & 1.14 \\
	\hline
	17 & 16 & 8 & 8 & 1.06 & 1 \\
	\hline	
	18 & 16 & 9 & 8 & 1.13 & 1.13 \\
	\hline
	19 & 17 & 9 & 8 & 1.12 & 1.13 \\
	\hline
	20 & 18 & 10 & 9 & 1.11 & 1.11 \\
	\hline
	21 & 19 & 10 & 9 & 1.11 & 1.11 \\
	\hline
	22 & 20 & 11 & 10 & 1.10 & 1.10 \\
	\hline
	23 & 21 & 11 & 10 & 1.10 & 1.10 \\
	\hline
	24 & 21 & 12 & 10 & 1.14 & 1.20 \\
	\hline
	25 & 22 & 12 & 11 & 1.14 & 1.09 \\
	\hline
	25 & 23 & 12 & 11 & 1.09 & 1.09 \\
	\hline
	26 & 23 & 13 & 11 & 1.13 & 1.18 \\
	\hline
	\cellcolor{red!25!blue!25} \tikzmark{b}{ \: 27 \: } & 24 & \cellcolor{red!25} \tikzmark{a}{ \: 13 \: } & 12 & 1.13 & 1.08 \\
	\hline
	28 & 25 & 14 & 12 & 1.12 & 1.17 \\
	\hline
	29 & 26 & 14 & 13 & 1.12 & 1.08 \\
	\hline
	30 & \cellcolor{red!45!yellow!45} \tikzmark{c}{ \: 27 \: } & \cellcolor{green!20} 15 & \cellcolor{yellow!45} \tikzmark{e}{ \: 13 \: }  & 1.11 & 1.15  \\ \cline{2-3}
	\hline
	31 & 28 & 15 & 14 & 1.11 & 1.07 \\
	\hline
	32 & 28 & 16 & 14 & 1.14 & 1.14 \\
	\hline
	33 & 29 & 16 & 14 & 1.14 & 1.14 \\
	\hline
\end{tabular}
\linka{a}{e}
\linkb{b}{a}
\linkc{c}{b}
\end{table}

\clearpage

\begin{table}[H]
\centering
	\caption{Clock tick associations, length associations, Lorentz factors for $v=c$.  To provide as much data as possible, we employ ever increasing abbreviated methods of displaying the data in later rows.}\label{table:Table2}
\begin{minipage}[t]{0.45\textwidth}\vspace{0pt}\hspace{-1pt}
\setlength{\tabcolsep}{4pt}
\begin{tabular}{|c|c|c|}
	\hline
	RF1 & RF2 & $\gamma_n = \gamma_m$\\ \cline{3-3} 
	Tick $n$ & Tick $n'$ & \multirow{2}{*}{$\frac{n}{n'}=\frac{m}{m'}$} \\ 
	Length $m$ & Length $m'$ & \\ 
	\hline
	\rowcolor{LightGray}
	1 & 1 & 1 \\
	\hline
	2 & 2 & 1 \\
	\hline
	3 & 2 & 1.5 \\
	\hline
	4 & 2 & 2 \\
	\hline
	\rowcolor{LightGray}
	5 & 3 & 1.67 \\
	\hline
	\rowcolor{LightGray}
	6 & 3 & 2 \\
	\hline
	\rowcolor{LightGray}
	7 & 3 & 2.33 \\
	\hline
	8 & 4 & 2 \\
	\hline
	9 & 4 & 2.25 \\
	\hline
	10 & 4 & 2.5 \\
	\hline
	11 & 4 & 2.75 \\
	\hline
	12 & 4 & 3 \\
	\hline
	\rowcolor{LightGray}
	13 & 5 & 2.6 \\
	\hline
	\rowcolor{LightGray}
	14 & 5 & 2.8 \\
	\hline
	\rowcolor{LightGray}
	15 & 5 & 3 \\
	\hline
	\rowcolor{LightGray}
	16 & 5 & 3.2 \\
	\hline
	\rowcolor{LightGray}
	17 & 5 & 3.4 \\
	\hline
	\hline
	$n_6=18$ & \multirow{2}{*}{\raisebox{-4pt}{$n'_6=6$}} & $\frac{n=18}{n'=6}=3$ \\
	\raisebox{4pt}{$\downarrow$} &   & \vdots \\
	\hline \rowcolor{LightGray}
	$n_7=25$ &  & 3.57 \\
	\rowcolor{LightGray}
	\raisebox{4pt}{$\downarrow$} & \multirow{-2}{*}{$n'_7=7$} & \vdots  \\ 
	\hline
	32 & \multirow{2}{*}{\raisebox{-4pt}{8}} & 4 \\
	\raisebox{4pt}{$\downarrow$} &  & \vdots  \\ 
	\hline \rowcolor{LightGray}
	41 &  & 4.56 \\
	\rowcolor{LightGray}
	\raisebox{4pt}{$\downarrow$} & \multirow{-2}{*}{9}  & \vdots  \\ 
	\hline
	50 & \multirow{2}{*}{\raisebox{-4pt}{10}} & 5 \\
	\raisebox{4pt}{$\downarrow$} &  & \vdots  \\ 
	\hline \rowcolor{LightGray}
	61 &  & 5.55 \\
	\rowcolor{LightGray}
	\raisebox{4pt}{$\downarrow$} & \multirow{-2}{*}{11} & \vdots  \\ 
	\hline
	72 & \multirow{2}{*}{\raisebox{-4pt}{12}} & 6 \\
	\raisebox{4pt}{$\downarrow$} &  & \vdots  \\ 
	\hline
	\end{tabular}
	\end{minipage} \hfill
	\begin{minipage}[t]{0.52\textwidth}\vspace{0pt}
	\setlength{\tabcolsep}{4pt}
	\begin{tabular}{|c|c|c|}
	\hline
	RF1 & RF2 & $\gamma_n = \gamma_m$\\ \cline{3-3} 
	Tick $n$ & Tick $n'$ & \multirow{2}{*}{$\frac{n}{n'}=\frac{m}{m'}$} \\ 
	Length $m$ & Length $m'$ & \\ 
	\hline \rowcolor{LightGray}
	85 &  & 6.54 \\
	\rowcolor{LightGray}
	\raisebox{4pt}{$\downarrow$} & \multirow{-2}{*}{13} & \vdots  \\ 
	\hline
	98 & \multirow{2}{*}{\raisebox{-4pt}{14}} & 7 \\
	\raisebox{4pt}{$\downarrow$} &  & \vdots  \\ \hline 
	\hline \rowcolor{LightGray} 
	113 $\downarrow$ & 15 & 7.53, $\dots$ \\
	\hline
	128 $\downarrow$ & 16 & 8, $\dots$ \\
	\hline
	\rowcolor{LightGray} 
	145 $\downarrow$ & 17 & 8.53, $\dots$ \\
	\hline
	162 $\downarrow$ & 18 & 9, $\dots$ \\
	\hline \rowcolor{LightGray} 
	181 $\downarrow$ & 19 & 9.52, $\dots$ \\
	\hline
	200 $\downarrow$ & 20 & 10, $\dots$ \\
	\hline \rowcolor{LightGray} 
	221 $\downarrow$ & 21 & 10.52, $\dots$ \\
	\hline
	242 $\downarrow$ & 22 & 11, $\dots$ \\
	\hline \rowcolor{LightGray} 
	265 $\downarrow$ & 23 & 11.52, $\dots$ \\
	\hline
	288 $\downarrow$ & 24 & 12, $\dots$ \\
	\hline \rowcolor{LightGray} 
	313 $\downarrow$ & 25 & 12.52, $\dots$ \\
	\hline
	338 $\downarrow$ & 26 & 13, $\dots$ \\
	\hline \rowcolor{LightGray} 
	365 $\downarrow$ & 27 & 13.52, $\dots$ \\
	\hline
	392 $\downarrow$ & 28 & 14, $\dots$ \\
	\hline \rowcolor{LightGray} 
	421 $\downarrow$ & 29 & 14.52, $\dots$ \\
	\hline
	450 $\downarrow$ & 30 & 15, $\dots$ \\
	\hline \rowcolor{LightGray} 
	481 $\downarrow$ & 31 & 15.52, $\dots$ \\
	\hline
	512 $\downarrow$ & 32 & 16, $\dots$ \\
	\hline \rowcolor{LightGray} 
	545 $\downarrow$ & 33 & 16.52, $\dots$ \\
	\hline
	578 $\downarrow$ & 34 & 17, $\dots$ \\
	\hline \rowcolor{LightGray} 
	613 $\downarrow$ & 35 & 17.51, $\dots$ \\
	\hline
	648 $\downarrow$ & 36 & 18, $\dots$ \\
	\hline \rowcolor{LightGray} 
	685 $\downarrow$ & 37 & 18.51, $\dots$ \\
	\hline
	722 $\downarrow$ & 38 & 19, $\dots$ \\
	\hline \rowcolor{LightGray} 
	761 $\downarrow$ & 39 & 19.51, $\dots$ \\
	\hline
	800 $\downarrow$ & 40 & 20, $\dots$ \\
	\hline \rowcolor{LightGray} 
	841 $\downarrow$ & 41 & 20.51, $\dots$ \\
	\hline
	882 $\downarrow$ & 42 & 21, $\dots$ \\
	\hline \rowcolor{LightGray} 
	925 $\downarrow$ & $43(\approx \sqrt{2n})$ & $21.51(\approx \sqrt{\frac{n}{2}})$ \\
	\hline
\end{tabular}
\end{minipage}
\end{table}

\clearpage

\section{\label{sec:Space_Transformation}The Lorentz Space Transformation}
The derivation of the Lorentz space transformation equations for DST is similar to the procedure used for CST.  For the reader's convenience, the CST forms of the equations are given below. 
\begin{subequations}
\begin{align}
&x = \gamma \left ( x'+ vt' \right ) \label{CST_Space_LorentzA}\\
&x' = \gamma \left ( x - vt \right ) \label{CST_Space_LorentzB}
\end{align}
\end{subequations}
\vspace{-0.5cm}
\begin{figure}[H]
	\centering\includegraphics[width=9cm]{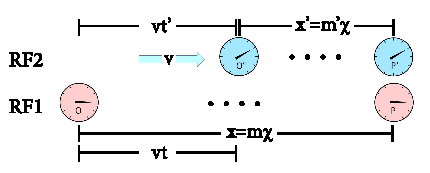}
	\caption{The four clocks used to derive Eq. \eqref{Lorentz_Space_A}.  The procedure for DST is not more difficult than it is for CST, but care must be given to the appropriate choice for $\gamma$ because it depends on both velocity and distance in DST.}
	\label{fig:Fig5}
\end{figure}

The only complication that arises when working in DST is how to appropriately contract lengths in DST.  Shown in Fig. \ref{fig:Fig5} is the typical diagram used for the derivation. Each reference frame has one particle at its origin and another particle at some distance away from its origin.  Reference frame RF2 is moving to the right with a velocity $v$ relative to RF1.  We must first transform from the distance $m\chi$ in RF1 to the corresponding distance in RF2.  From RF2's perspective, the distance $m\chi$ is a moving distance, hence $P'$ will have traveled a distance $\gamma^{-1}_m m\chi$ when it overlaps with the particle $P$.  Once this distance is transformed to the appropriate distance in RF2, the addition (and subtraction) of lengths \textit{within a single reference frame (RF2 in this case)} can be done in the normal way regardless of space and time being discrete or continuous.    We arrive at the following equation:
\begin{equation}\label{Lorentz_Space_A}
 \gamma^{-1}_m m\chi = m'\chi+vt'
\end{equation}
\noindent or in terms of normalized values:
\begin{equation}\label{Lorentz_Space_B}
 \gamma^{-1}_m m = m'+ p'= m'+ \Bigl\lfloor \frac{v_{avg}}{c} n' \Big\rfloor
\end{equation}
\noindent where $p'$ is equal to the normalized (relative to $\chi$) distance traveled by $O'$ and $P'$ over a time $t'$, which can also be expressed as $p'= \lfloor \frac{v_{avg}}{c}n' \rfloor$.  The second Lorentz space transformation is then easily derived:
\begin{equation}\label{Lorentz_Space_C}
 \gamma^{-1}_{m'}m' = m-p = m - \Bigl\lfloor \frac{v_{avg}}{c} n \Big\rfloor
\end{equation}
\noindent where $p$ is equal to the normalized (relative to $\chi$) distance traveled by $O$ and $P$ over a time $t$, namely $p = \lfloor \frac{v_{avg}}{c}n \rfloor$.

\section{\label{sec:Clock_Syncrohnization}Clock Synchronization}
\subsection{\label{sec:CST_Clock_Sync} Clock Synchronization in CST}
First, let us review clock synchronization in CST by simply manipulating the two Lorentz space transformations given by Eqs.  \eqref{CST_Space_LorentzA} and \eqref{CST_Space_LorentzB}.  In these equations, we solve for $ct'$ ant $ct$ in terms of the coordinates in the other reference frame:
\begin{subequations}
\begin{align}
&ct = \gamma \left ( ct' + \frac{v}{c}x' \right ) \label{CST_Time_Lorentz_A} \\
&ct' = \gamma \left ( ct - \frac{v}{c}x \right )  \label{CST_Time_Lorentz_B}
\end{align}
\end{subequations}
\noindent which can be put into the form:
\begin{subequations}\label{CST_Time_Lorentz_E}
\begin{align}
&\gamma^{-1}ct = ct' + \frac{v}{c}x' \label{CST_Time_Lorentz_C} \\
&\gamma^{-1}ct' =  ct - \frac{v}{c}x  \label{CST_Time_Lorentz_D}
\end{align}
\end{subequations}

The roles of the terms in Eqs. \eqref{CST_Time_Lorentz_C} and \eqref{CST_Time_Lorentz_D} can be readily identified.  Equation \eqref{CST_Time_Lorentz_C} is used for the case where an observer $O'$ in RF2 recorded the time and position of an event as $(x',t')$.  If the event happened right \textit{at} $O'$, i.e., namely, at $x'=0$, then the observer $O$ in $RF1$ would know that the value $t'$ is simply the contraction of the time at which he records the event as occurring---this fact is a simple application of the time dilation.  However, there is the second term on the right side of Eq. \eqref{CST_Time_Lorentz_C}, namely $+vx'/c$.  This term is not present in nonrelativistic mechanics and is a correction factor---an amount of time that needs to be \textit{added} to the time of an event at position $x'$ recorded by $O'$ such that, upon subsequent dilation of this shifted time (i.e., multiplication by $\gamma$), one obtains the correct time of the event as recorded by $O$ in RF1.  Likewise for the terms in Eq. \eqref{CST_Time_Lorentz_D}, the amount of time $vx/c$ is required to be \textit{subtracted} from the time of an event at position $x$ recorded by  $O$ such that, upon subsequent dilation of this shifted time, one obtains the correct time of the event as recorded by $O'$.  Thus the terms $vx'/c$ in Eq. \eqref{CST_Time_Lorentz_C} and $-vx/c$ in Eq. \eqref{CST_Time_Lorentz_D} are the clock asychronicity terms (CATs)---the amount of time that observers in the \textit{other} reference frame believe a clock is asynchronous with respect to the clock at the origin of that other reference frame.  

At this point, one may be tempted to jump directly to what may appear to be the obvious DST versions of Eqs. \eqref{CST_Time_Lorentz_C} and \eqref{CST_Time_Lorentz_D} (with $t=n\tau$, $x=m\chi$, $t'=n'\tau$, $x'=m'\chi$):
\begin{subequations}\label{CST_Time_Lorentz_E}
\begin{align}
&\gamma_n^{-1}n = n' + \Big\lfloor \frac{v_{avg}}{c}m' \Big\rfloor \label{CST_Time_Lorentz_F} \\
&\gamma_{n'}^{-1}n' =  n - \Big\lfloor \frac{v_{avg}}{c}m\Big\rfloor  \label{CST_Time_Lorentz_G}
\end{align}
\end{subequations}
However, the choice of the floor operation in Eqs. \eqref{CST_Time_Lorentz_F} and \eqref{CST_Time_Lorentz_G} rather than the ceiling operation may strike some readers as arbitrary.  The choice of the floor operation in the Lorentz \textit{space} transformation (Eq. \eqref{Lorentz_Space_B} and \eqref{Lorentz_Space_C}) is easily justified since it was based on our choice of when movement of a particle occurs over some interval.  \textit{(Does a traveling particle with $v_{avg}=c/10$ translate a hodon at the first tick of a each ten hodon duration or the tenth tick?  We make the arbitrary choice of the tenth tick.)}\footnote{The DST CAT terms $\lfloor \frac{v_{avg}}{c}m' \rfloor$ and $\lfloor \frac{v_{avg}}{c}m \rfloor$ in the Lorentz \textit{time} transformation (Eq. \eqref{CST_Time_Lorentz_F} and \eqref{CST_Time_Lorentz_G}) have similar forms relative to their space counterparts because their cause is similar:  a movement of an origin of the reference frame relative to the other reference frame.  In the Lorentz \textit{space} transformation equations it is the movement of the \textit{spatial origin} (i.e., the point in spacetime where $x'=0$ as a function of time), whereas with the Lorentz \textit{time} transformation equations it is the movement of the \textit{time origin} (the point in spacetime where $t'=0$ as a function of space).}  To more adequately justify the forms of the time transformation equations given by Eqs. \eqref{CST_Time_Lorentz_F} and \eqref{CST_Time_Lorentz_G}, let us employ a second well-known method to assess clock asynchronicity...one that uses photons to synchronize clocks.  Being so well-known, let us skip to photon-mediated clock synchronization in DST; any reader that needs a review of the procedure in CST can consult any introductory book on special relativity (e.g., \cite{Helliwell2010}).

\subsection{\label{sec:Clock_Syncrohnization_DST}Clock Asynchronization in DST}
\begin{figure}[H]
	\centering\includegraphics[width=11cm]{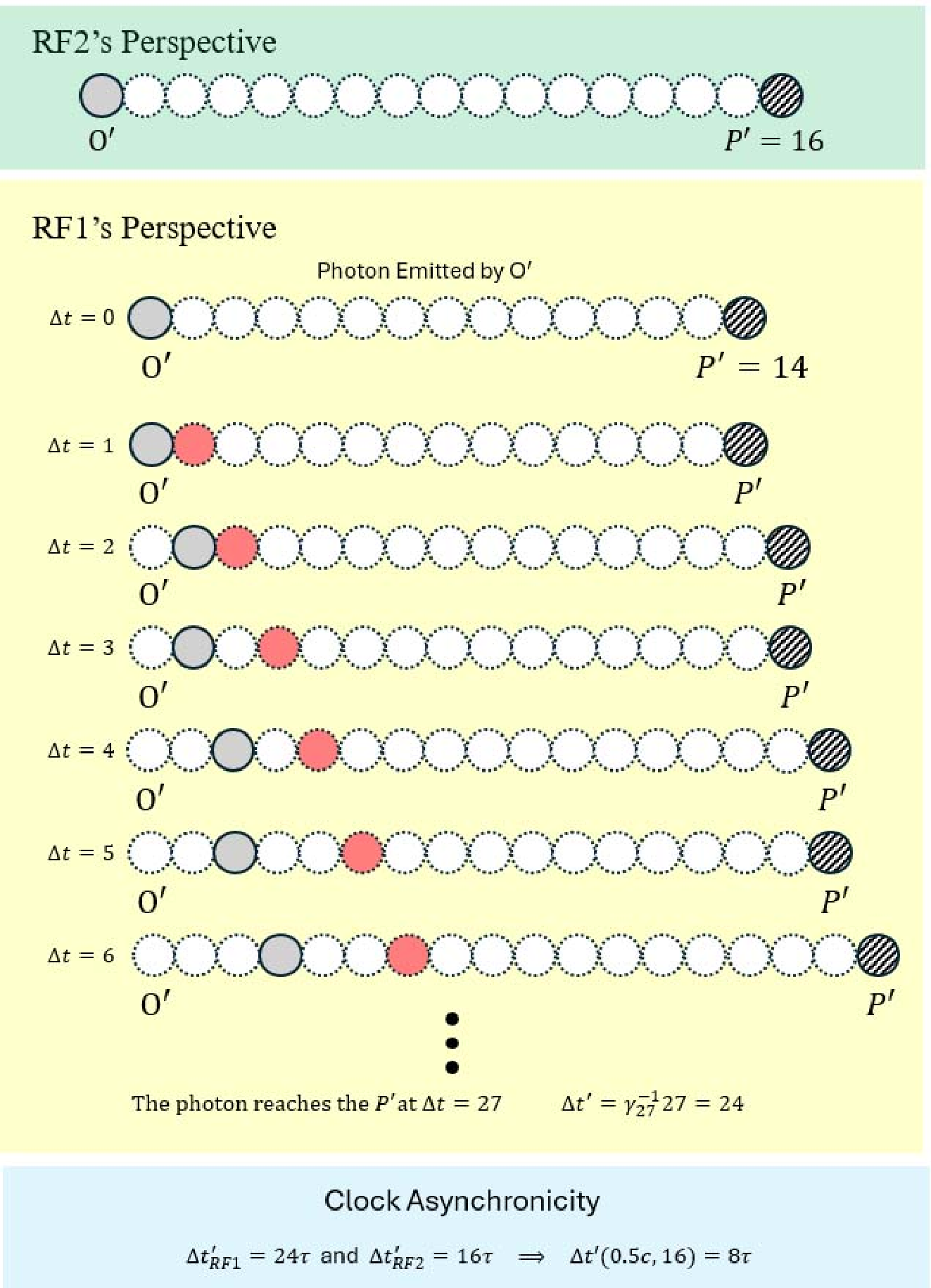}
	\caption{The process of synchronizing clocks in a reference frame with $v_{avg}=0.5c$.  Performing the process for discrete spacetime is similar to the process used for CST.  }
	\label{fig:Fig6}
\end{figure}

Photon-mediated clock synchronization in DST is very similar to the procedure in CST.  Consider two clocks in RF2 that are separated by a distance $m'$ and both moving with an average velocity $v_{avg}$ to right relative to RF1, as shown for an example case of $v_{avg}=0.5c$ in Fig. \ref{fig:Fig6}.  RF1 observers will observe the moving distance $m'$ to be contracted to $\gamma^{-1}_{m'}m'$.  This contracted distance and the extra distance traveled by $P'$ (denoted as $v_{avg}t_{expected}$) together needs to be covered by the photon, hence:\footnote{The quantity $v_{avg}t_{expected}$ needs to be an integer multiple of $\chi$ as explained later in this section.  In the method described in this section, we ultimately never use $v_{avg}t_{expected}$ but rather use the tabulated value of $p(n)$ or the equation for $p$ in terms of ticks $n$.}
\begin{equation}\label{General_Time_Asy_A}
	\gamma^{-1}_{m'}m'\chi + v_{avg}t_{expected} = ct_{expected}
\end{equation}
\noindent or:
\begin{equation}\label{General_Time_Asy_B}
	\gamma^{-1}_{m'}m' = n - p(n)
\end{equation}
\noindent where $n=t_{expected}/\tau$, and with $\lfloor \frac{v_{avg}}{c}n \rfloor=p(n)$ being the normalized distance (relative to $\chi$) traveled by \textit{Clock P'} during the $n$ ticks of RF1's clocks.  Once $t_{expected}$ is found, it is contracted in the appropriate way:
\begin{equation}\label{General_Time_Asy_C}
	t'_{expected} = \gamma^{-1}_{t_{expected}}t_{expected}
\end{equation}
\noindent Then one subtracts from this value of $t'_{expected}$ the actual time recorded by \textit{Clock P'}, namely $t'_{actual}=d/c=m'\chi/c=m'\tau$ to obtain the CAT:
\begin{equation}\label{General_Time_Asy_D}
	\Delta t'(v_{avg},m') = t'_{expected} - m'\tau
\end{equation}
Once the CAT is calculated, one can easily derive the Lorentz time transformation for DST:
\begin{equation}\label{Lorentz_Time_B}
\gamma^{-1}_{t}t = t' + \Delta t'(v_{avg},m')
\end{equation}

\noindent In a similar way, we can obtain the CAT for RF1 ($\Delta t (v_{avg},m)$):
\begin{equation}\label{General_Time_Asy_E}
	\gamma^{-1}_{m} m = n' + p'(n')
\end{equation}
\noindent where $n'=t'_{expected}/\tau$,  and with $\lfloor \frac{v_{avg}}{c}n' \rfloor=p'(n')$ being the normalized distance (relative to $\chi$) traveled by \textit{Clock P} during the $n'$ ticks of RF2's clocks.  We then have:
\begin{equation}\label{General_Time_Asy_F}
	t_{expected} = \gamma^{-1}_{t'_{expected}}t'_{expected}
\end{equation}
\begin{equation}\label{General_Time_Asy_G}
	\Delta t(v_{avg},m) = t_{expected} - m\tau
\end{equation}

\begin{figure}[H]
	\centering\includegraphics[width=11cm]{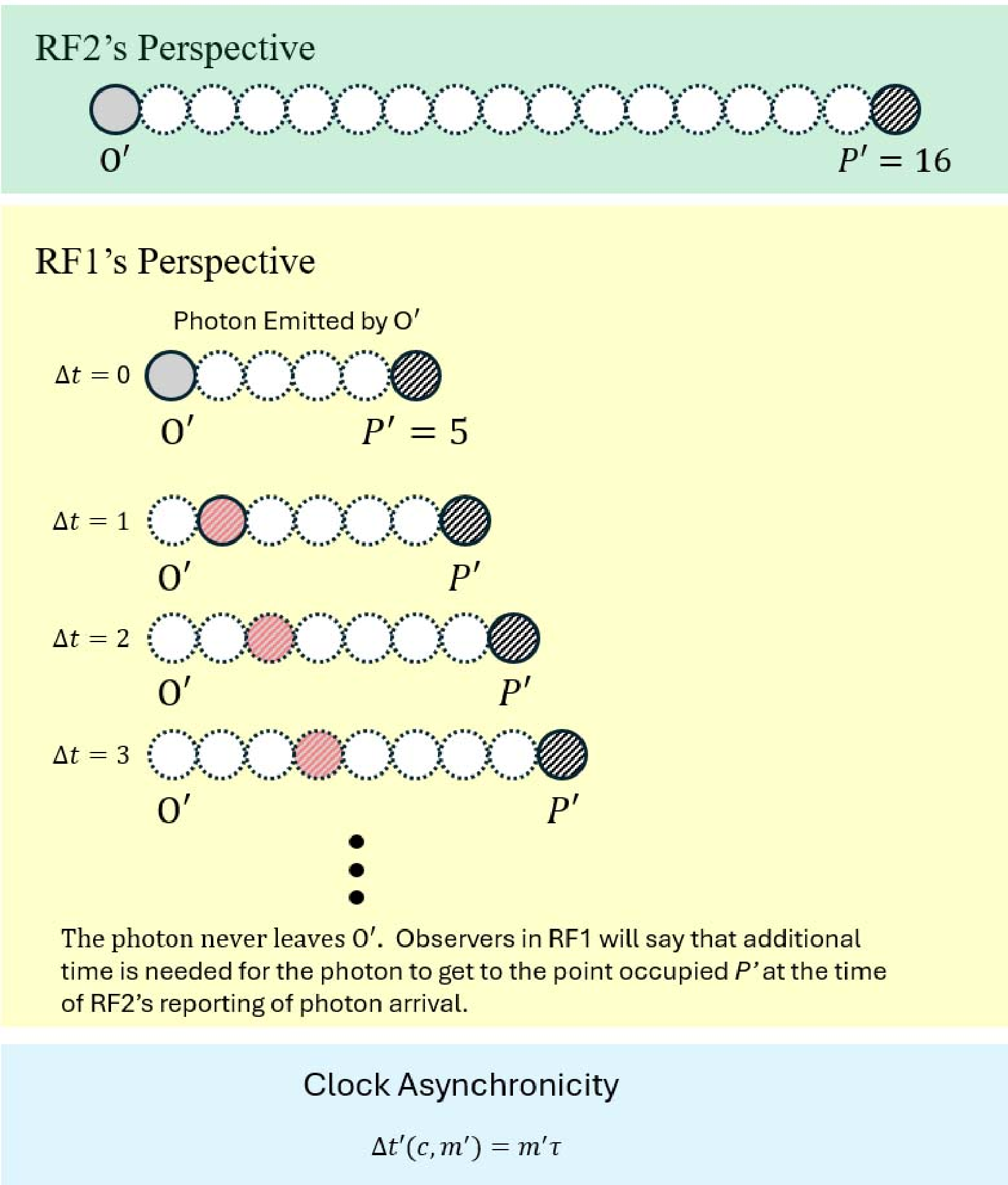}
	\caption{If the reference frame is traveling at $v=c$, then, from RF1's perspective, the photon never propagates away from $O'$. }
	\label{fig:Fig7}
\end{figure}

Consider the example calculation of $\Delta t'(v_{avg}=0.5c,m'=16)$ shown in Fig. \ref{fig:Fig6} where $v_{avg}=0.5c$ and for a distance $m'=16$ between \textit{Clock O'} and \textit{Clock P'}.  We get the necessary positions and contractions from Table 1, which we collect below in Tables \ref{table:Table5}-\ref{table:Table7} in ways more convenient for the reader.   We see from the Table \ref{table:Table5} that $m'=16$ is contracted to $\gamma^{-1}_{16}16=14$.  We use this value of $14$ in Eq. \eqref{General_Time_Asy_B} and data from Table \ref{table:Table6} to get $14=n-p$ with $n=27$ and $p=13$.  We then use this value for $n(=27)$ and find its contraction from Table \ref{table:Table7} to be $\gamma^{-1}_{27}27=24$.  Thus $t'_{expected}=24\tau$, whereas $t_{actual}=16\tau$.  Finally, using Eq. \eqref{General_Time_Asy_D} we get $\Delta t'(v=0.5c,m'=16) = 24\tau - 16\tau= 8\tau$.\footnote{As with all other processes described in this work, simple computer programs can be written that perform all these calculations and associations or obtained by contacting the author.}  This value of $8\tau$ is what is expected if one naively uses the CST version of the time asynchronicity equation, namely, $\Delta t' (v=0.5c,x'=16\chi)=vx'/c^2=0.5\cdot 16\tau=8\tau$.  While more complicated, the use of the proper procedure developed in this paper to obtain time asynchronicity (along with everything else) is important to instill confidence in the results obtained and buttress the validity of the DST model.  
\begin{table}[H]
	\caption{Length Contraction Associations for $v_{avg}=0.5c$}\label{table:Table5}\vspace{-.2cm}
	\noindent \begin{tabular}{|c||c|c|c|c|c|c|c|c|c|c|c|c|c|c|}
		\hline
		$m'$ & 0 & 1 & 2 & 3 & 4 & 5 & 6 & 7 & 8 & 9 & 10 & 11 & 12 & 13 \\
		\hline
		$\gamma^{-1} m'$ & 0 & 1 & 2 & 3 & 4 & 5 & 6 & 6 & 7 & 8 & 9 & 10 & 11 & 12\\
		\hline
	\end{tabular}
	...\\
	\\
	...
	\begin{tabular}{|c|c|c|c|c|c|c|c|c|c|c|c|c|c|c|c|}
		\hline
		14 & 15 & 16 & 17 & 18 & 19 & 20 & 21 & 22 & 23 & 24 & 25 & 26 & 27 & 28 \\
		\hline
		12 & 13 & 14 & 15 & 16 & 17 & 18 & 19 & 20 & 20 & 21 & 22 & 23 & 24 & 25 \\
		\hline
	\end{tabular}
\end{table}
\begin{table}[H]
\caption{Time ($n$) and Associated Travel Distance ($p$) for $v_{avg}=0.5c$}\label{table:Table6}\vspace{-.2cm}
\noindent \begin{tabular}{|c||c|c|c|c|c|c|c|c|c|c|c|c|c|c|c|}
		\hline
		$n$ & 1 & 2 & 3 & 4 & 5 & 6 & 7 & 8 & 9 & 10 & 11 & 12 & 13 & 14 &15\\
		\hline
		$p$ & 0 & 1 & 1 & 2 & 2 & 3 & 3 & 4 & 4 & 5 & 5 & 6 & 6 & 7 & 7\\
		\hline
\end{tabular}
...\\
\\
...
\begin{tabular}{|c|c|c|c|c|c|c|c|c|c|c|c|c|c|c|}
	\hline
		16 & 17 & 18 & 19 & 20 & 21 & 22 & 23 & 24 & 25 & 26 & 27 & 28 & 29 & 30 \\
	\hline
		8 & 8 & 9 & 9 & 10 & 10 & 11 & 11 & 12 & 12 & 13 & 13 & 14 & 14 & 15 \\
	\hline
\end{tabular}
\end{table}
\begin{table}[H]
	\caption{Time Dilation Associations for $v_{avg}=0.5c$}\label{table:Table7}\vspace{-.2cm}
	\noindent \begin{tabular}{|c||c|c|c|c|c|c|c|c|c|c|c|c|c|c|}
		\hline
		$n'$ & 1 & 2 & 3 & 4 & 5 & 6 & 7 & 8 & 8 & 9 & 10 & 11 & 12 & 13  \\
		\hline
		$\gamma^{-1} n'$ & 1 & 2 & 3 & 4 & 5 & 6 & 7 & 7 & 8 & 9 & 9 & 10 & 11 & 12 \\
		\hline
	\end{tabular}
	...\\
	\\ 
	...
	\begin{tabular}{|c|c|c|c|c|c|c|c|c|c|c|c|c|c|c|c|}
		\hline
		14 & 15 & 16 & 17 & 18 & 19 & 20 & 21 & 22 & 23 & 24 & 25 & 26 & 27 & 28 \\
		\hline
		13 & 14 & 14 & 15 & 16 & 17 & 18 & 19 & 20 & 21 & 21 & 22 & 23 & 24 & 25 \\
		\hline
	\end{tabular}
\end{table}
The above method of synchronizing clocks works well for all relative velocities except for $v=c$; this case requires special attention.  Figure \ref{fig:Fig7} is similar to Fig.  \ref{fig:Fig6} but shows the case with $v=c$.  What we see is that, from the perspective of RF1, not only does the photon never reach $P'$, it never advances beyond $O'$!  Consider the very instant RF2 observers report that the photon arrives at $P'$ at $t'=d'/c$,  RF1 observers will say that the experiment had not even started(!)...and that the photon would require an additional  time of $t'=d'/c$ (beyond the time reported by RF2 observers) to arrive at that exact position in RF2.    Therefore, RF1 believes that $O'$'s clock should be adjusted forward by $d'/c$ before contracting it to obtain the correct time in RF1's frame of reference.  Hence, the clock asynchronicity factor for a $O'$-$P'$ separation of $m'$ is:
\begin{equation}\label{General_Time_Asy_H}
	\Delta t'(c,m') = m'\tau 
\end{equation}
\noindent The inverse is easily seen to be:
\begin{equation}\label{General_Time_Asy_I}
	\Delta t(c,m) = -m\tau 
\end{equation}
We therefore have the two following Lorentz time transformation equations for reference frames with a \textit{light-speed relative velocity}:
\begin{subequations}
	\begin{align}
		&\gamma^{-1}_{n'}n' =  n - m    \label{Lorentz_Time_G} \\
		&\gamma^{-1}_{n}n =  n' + m'   \label{Lorentz_Time_H}
	\end{align}
\end{subequations}
\noindent
This is in agreement with what is obtained in CST as we let $v \rightarrow c$, the only difference is that the Lorentz factors $\gamma_{n}$ and $\gamma_{n'}$ \textit{remain finite} in DST ($\gamma^{-1}_{n'}$ and $\gamma^{-1}_{n}$ are nonzero).  
After having completed clock synchronization using the laborious procedure of photon synchronization only to have the results yield what is obtained using the term $\Delta t(v_{avg},m))=\lfloor \frac{v_{avg}}{c}m \rfloor \tau$ makes a revisit of Eqs. \eqref{CST_Time_Lorentz_F} and \eqref{CST_Time_Lorentz_G} important. One reason is that more important physics and philosophical foundations may be gleaned from the equations once analyzed more thoroughly.  A second reason is that it provides a much easier way to calculate $\Delta t(v_{avg},m)$ and $\Delta t'(v_{avg},m')$.  When studying Fig. \ref{fig:Fig7} and the term $\lfloor \frac{v_{avg}}{c}m \rfloor$, we see that the term gives the number of ticks in RF2 that RF1 observers will say that both the \textit{Clock O'} and the photon have moved by one hodon---this tick would not have been counted by RF2 observers but should be added to the time $n'\tau$ recorded by RF2 observers  to obtain, upon dilation, the correct time in RF1.  This simple physical explanation, namely accounting for the erroneously skipped ticks by $RF2$ in its accounting of time, is very satisfactory...more so than the explanation of clock asynchronicity in CST.

\section{\label{sec:Lorentz_DST}The Lorentz Transformation Equations for DST}
Collecting the results from the prior sections, we have the normalized LTEs for DST as:
\begin{subequations} \label{Lorentz_Final_DST}
	\begin{align}
		\gamma^{-1}_m m &= m' + \Delta (v_{avg},n')  \label{Lorentz_Final_DST_A} \\
		\gamma^{-1}_n n &= n' + \Delta (v_{avg},m')   \label{Lorentz_Final_DST_B} \\
		\gamma^{-1}_{m'} m' &= m + \Delta (v_{avg},n) \label{Lorentz_Final_DST_C}  \\
		\gamma^{-1}_{n'} n' &= n + \Delta (v_{avg},m)  \label{Lorentz_Final_DST_D}
	\end{align}
\end{subequations}

\noindent with $x = m\chi$, $t = n\tau$, $x'=m'\chi$, and $t'=n'\tau$, and with the terms $\Delta (v_{avg},j)$ and $\gamma^{-1}_j j$ given by:
\begin{equation}\label{GNAST}
\Delta (v_{avg},j) =\pm \Big \lfloor \frac{v_{avg}}{c} j \Big \rfloor
\end{equation}
\begin{equation}
	\gamma^{-1}_j j = j_{c} = \Biggl \lceil   \sqrt{j^2 - \Big\lfloor \frac{v_{avg}}{c}j \Bigr\rfloor^2 +2j+1} -1  \Biggr\rceil  \label{gamma_expression_DST_a}
\end{equation}

\noindent with $j_c$ is the contracted quantity, $j=n',m',m,n$, and with $+/-$ chosen as appropriate for the reference frame:  $+$ for RF2 coordinates, $-$ for RF1 coordinates.  The above equations apply equally well for light-speed reference frames ($v_{avg}=c$) as they do for sub light-speed reference frames ($v_{avg}<c$).  Except for light-speed systems, a closed-form equation to dilate a quantity cannot be derived as readily as that for a quantity's contraction (i.e., Eq. \eqref{gamma_expression_DST_a}).  However, such an operation will prove useful when we study velocity addition in the next section and any other derivation that needs contraction of a quantiy.  We therefore express such a dilation operation as:
\begin{equation}\label{gamma_expression_DST_b}
	\gamma_{j_c} j_c = j
\end{equation}
\noindent The lack of a closed-form expression for $\gamma_{j_c} j_c$, however, is not too much of a problem because one can readily use Eq. \eqref{gamma_expression_DST_a} to obtain the $(j,j_c)$ associations.

\section{\label{sec:Velocity_Addition}Velocity Addition in Discrete Spacetime}
It is important to develop the equation for the addition of velocities specified in two different reference frames in DST.  To develop the appropriate equation, let us have the moving reference frame's (RF2) average velocity relative to the rest reference frame RF1 as $v_{2,1}$.  Also, let a particle in RF2 have an average velocity $v_{p,2}$.    For CST, the relativistic velocity addition equation that provides the velocity of the particle in RF1 is well-known:
\begin{equation}
	v_{p,1} = \frac{v_{p,2} + v_{2,1}}{1+v_{p,2}v_{2,1}/c^2}
\end{equation}
\noindent To develop the equivalent equation appropriate for DST, we start with Eq. \eqref{Lorentz_Final_DST_A} and have $m_{p,2}'$ as the position of the particle in RF2 that starts at the origin of RF2 at $n'=0$ and is moving in RF2 according to the following equation:
\begin{equation}\label{mprime}
	m'_{p,2} = \Big \lfloor \frac{v_{p,2}}{c}n' \Big \rfloor
\end{equation}
\noindent We use Eq. \eqref{mprime} in Eq. \eqref{Lorentz_Final_DST_A} and \eqref{Lorentz_Final_DST_A}, which can then be put into the following form:
\begin{subequations}
	\begin{align}
		m_{p,1} &= \gamma_{\left ( \lfloor \frac{v_{p,2}}{c}n' \rfloor + \lfloor \frac{v_{2,1}}{c}n'  \rfloor \right )} \left (  \Big \lfloor \frac{v_{p,2}}{c}n' \Big \rfloor + \Big \lfloor \frac{v_{2,1}}{c}n' \Big \rfloor \right ) \\
		n_{p,1} &= \gamma_{ \left ( n' +  \lfloor \frac{v_{2,1}}{c}  \lfloor \frac{v_{p,2}}{c}n'  \rfloor  \rfloor \right )} \left ( n' + \Big \lfloor \frac{v_{2,1}}{c} \Big \lfloor \frac{v_{p,2}}{c}n' \Big \rfloor \Big \rfloor  \right )
	\end{align}
\end{subequations}
\noindent We then have the velocity of the particle in RF1 (i.e., $v_{p,1}$) as:
\begin{equation}
	v_{p,1} = \frac{m_{p,1}}{n_{p,1}}c = \frac{  \gamma_{\left ( \lfloor \frac{v_{p,2}}{c}n' \rfloor + \lfloor \frac{v_{2,1}}{c}n'  \rfloor \right )} \left (  \Big \lfloor \frac{v_{p,2}}{c}n' \Big \rfloor + \Big \lfloor \frac{v_{2,1}}{c}n' \Big \rfloor \right )}{  \gamma_{ \left ( n' +  \lfloor \frac{v_{2,1}}{c}  \lfloor \frac{v_{p,2}}{c}n'  \rfloor  \rfloor \right )} \left ( n' + \Big \lfloor \frac{v_{2,1}}{c} \Big \lfloor \frac{v_{p,2}}{c}n' \Big \rfloor \Big \rfloor  \right )} \label{DSTVelocityAdd}c
\end{equation}
\noindent where we have used the notation $\gamma_{j_c} j_c$ to denote the dilation operator $\gamma_{j_c}$ operating on $j_c$, as described in Section \ref{sec:Lorentz_DST}.

The system with $v_{p,2}=v_{2,1}=c$ will be important when we analyze Stadium in Section \ref{sec:Stadium}. In this case, Eq. \eqref{DSTVelocityAdd} becomes:
\begin{equation}
	v_{p,1} = \frac{m_{p,1}}{n_{p,1}}c  = \frac{\gamma_{\left (  n' + n' \right )} \left ( n' + n' \right )}{\gamma_{\left (  n' + n' \right )} \left ( n' + n' \right )}c = c \label{DSTVelocityAddc}
\end{equation}
\noindent We see that nature's speed-limit is maintained, the light-speed particle in RF2 is traveling at light-speed in RF1.

\section{\label{sec:Minkowski}Minkowski Spacetime for Discrete Spacetime}
\begin{figure}[H]
	\centering\includegraphics[width=10cm]{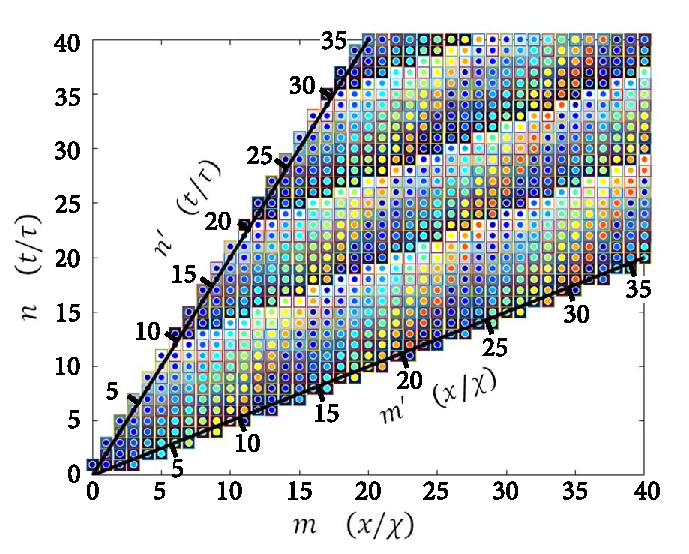}
	\caption{Minkowski space-time diagram for two reference frames, one at rest (RF1) and one (RF2) traveling at $v_{avg}=0.5c$ relative to RF1.  A colored colormap is used to indicate lines of constant $m'$, and a grayscale colormap is used to indicate lines of constant $n'$.  A consequence of space and time are discrete is that each $(m,n)$ point may be associated with multiple $(m',n')$ points.}
	\label{fig:Fig8}
\end{figure}

\begin{figure}[H]
	\centering\includegraphics[width=10cm]{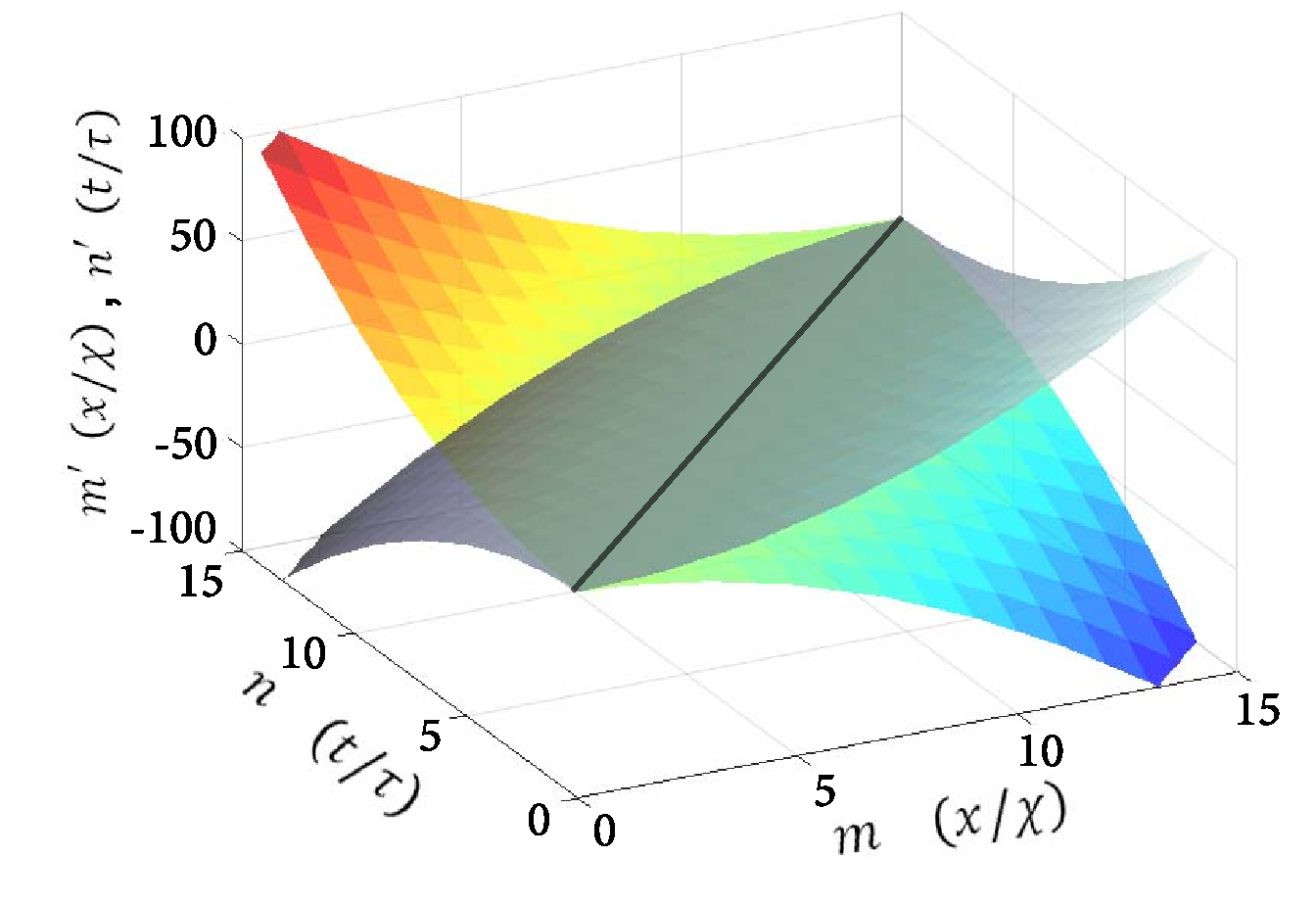}
	\caption{\textit{Light-Speed Motion}: A graph of $m'$ (grayscale) and $n'$ (shaded in colors) in the light-speed RF2 as a function of $m$ and $n$ in RF1.  In special relativity in CST, all space in the moving frame is compressed to the light-line (the black line).  However, in DST, space and time exists away from the light-line. }
	\label{fig:Fig9}
\end{figure}

With the Lorentz transformation equations (Eqs. \eqref{Lorentz_Final_DST_A}-\eqref{Lorentz_Final_DST_D} we are ready to construct a couple of examples of Minkowski space diagrams. An example of Minkowski spacetime diagrams for $v_{avg}=0.5c$ is shown in Fig. \ref{fig:Fig8} and one for $v=c$ is shown in Fig. \ref{fig:Fig9}.  Computer code to construct Minkowski DST diagrams for arbitrary $v$ is available by contacting the author.

We have now largely completed the list of key foundational concepts of special relativity in DST that were needed to be carefully analyzed: time dilation, length contraction, clock asynchronicity, Lorentz transformation equations, velocity addition, and Minkowski spacetime diagrams.  One final important item remains on our list of problems associated with DST and needs to be discussed:  Zeno's Stadium Paradox.

\section{\label{sec:Stadium} The Stadium Paradox}

\subsection{Bertrand Russell's Formulation of Zeno's Stadium Paradox}
Parmenides's and Zeno of Elea's paradoxes are some of the most amazing philosophical and mathematical thought experiments produced by the human mind.  A superficial reading of them often results in a sense of incredulity by most people who wonder why a small subset of people continue to study the paradoxes.  It is not only laypeople with this view; presumably, Diogenes himself seemed satisfied in refuting Zeno's paradoxes that aimed to prove that motion is impossible (specifically the Arrow Paradox) by simply standing up and walking!  However, the more one contemplates the paradoxes, the more intriguing they become. Such was the experience for the great Bertrand Russell who developed his theory of ``at-at motion'' in response to the paradoxes \citep[45]{Salmon2001}.   Russell stated that  ``the more one thinks about it (the Stadium Paradox), the more perplexing it becomes''.

\begin{figure}[H]
	\centering\includegraphics[width=8cm]{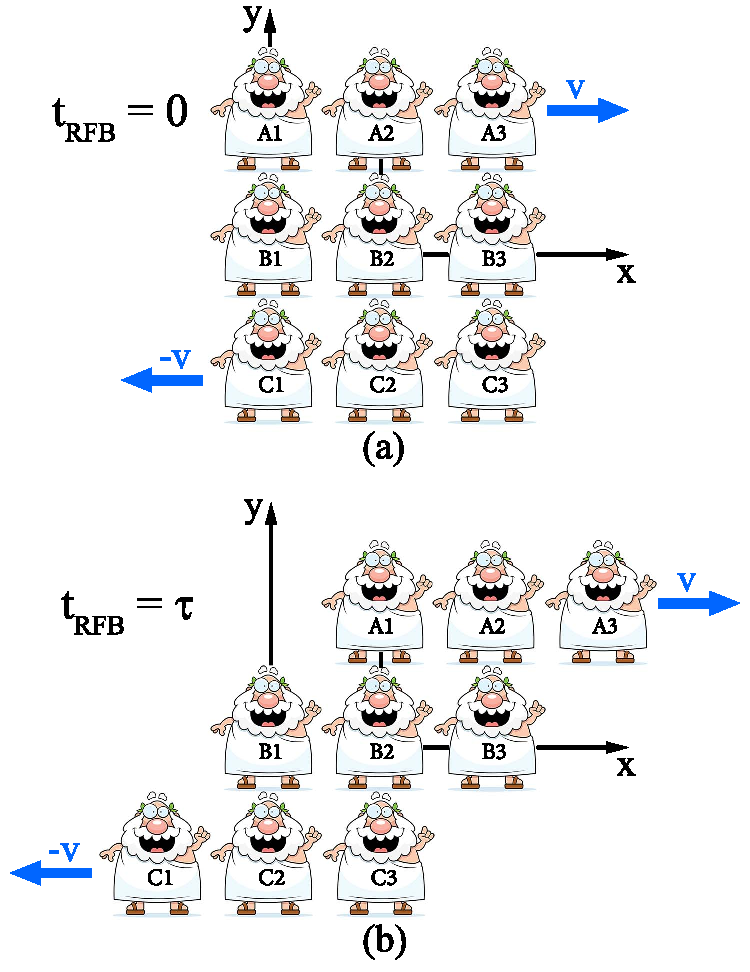}
	\caption{The Stadium Paradox as described by \citet{Salmon2001}, \citet{Whitrow1962}, \citet{Russell2003}, and \citet{Grunbaum1952}. (a) The initial configuration at $t=0$. (b) The configuration after one fundamental \textit{snapshot} in time, \textit{i.e.,} $t=1\tau$.  We see that $A1$ is over $C3$ at $t=1\tau$, leading to the question: at what time was $A1$ over $C2$?}
	\label{fig:Fig10}
\end{figure}

The Stadium Paradox is as follows: consider three rows of identical bodies in a stadium, rows $A$, $B$, and $C$, as shown in Fig. \ref{fig:Fig10}. For simplicity, let us consider three Athenians per row, labeled numerically left to right. Also assume that each body occupies the smallest unit of spatial distance, namely, a hodon $\chi$.  We set up the situation such that the Athenians in rows \textit{A} are traveling to the right at a speed of one hodon per chronon $\tau$ (the assumed smallest duration of time) relative to the Athenians in row B.  The Athenians in rows \textit{C} are traveling to the left at a speed of one hodon per chronon relative to the Athenians in row B.   We now look at two instances in time: $t=0$ and $t=1\tau$.  At $t=0$, $A1$ is directly above $B1$ which is directly above $C1$, and the same with the other two columns of Athenians.  At $t=\tau$, $A1$ has transitioned so that it is directly above $B2$, which is now directly above $C3$.  The troublesome question that naturally arises is:  when was $A1$ directly above $C2$?  The common answer is: presumably at $\tau/2$.  If so, the reality of $\tau/2$ must be accepted...therefore $\tau/2$ must be the smallest time duration, not $\tau$.  One would then perform this motion over again, but with $\tau/2$ as the duration of the chronon...alas, one sees that the reality of $\tau/4$ must be accepted!  One carries on this procedure \textit{ad infinitum}, at each iteration forcing the reality of a smaller fundamental time.  Thus, the paradox seems to support the position that there cannot be a fundamental \textit{smallest} temporal duration or spatial extent.   Before analyzing the paradox within the framework of DST let us see how modern conventional physics addresses the paradox, namely using special relativity in CST. 

\subsection{Relativistic Solution in Continuous Spacetime}
\begin{figure}[H]
	\centering\includegraphics[width=8cm]{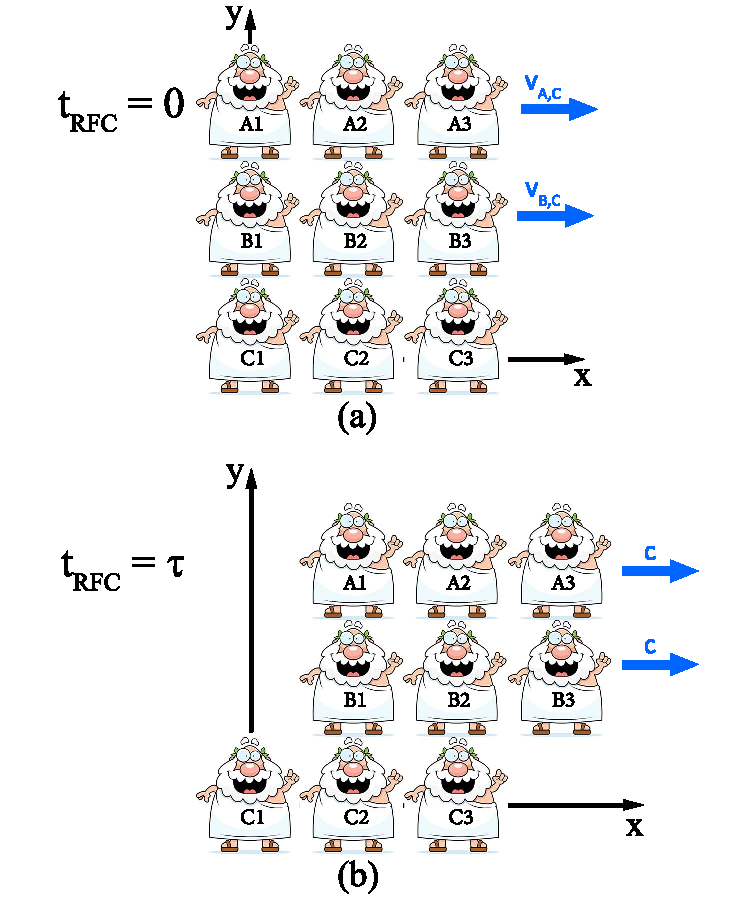}
	\caption{Using the laws of special relativity in CST, we find that the velocity of the \textit{Row A} Athenians relative to the \textit{Row C} Athenians (at rest in RFC) is $v_{A,C}=c$\dots the same as $v_{B,C}$. }
	\label{fig:Fig11}
\end{figure}

Assume that both \textit{Row A} and \textit{Row B} Athenians are in the same reference frame RFB, with \textit{Row B} Athenians at rest in RFB and with \textit{Row A} Athenians traveling at a velocity $v_{A,B}$ in RFB.  RFB is traveling at a velocity $v_{B,C}$ relative to the rest reference frame RFC.  \textit{Row C} Athenians are at rest in RFC.  In continuous spacetime, the relevant velocity transformation equation that provides the velocities of the \textit{Row A} Athenians relative to a rest reference frame RFC, namely $v_{A,C}$, is: 
\begin{equation}\label{CST_Velocity}
	v_{A,C}=\frac{v_{A,B} + v_{B,C}}{1 +v_{A,B}v_{B,C}/c^2}
\end{equation}
Now for the situation described in the paradox (Fig. \ref{fig:Fig10}), one sees that $v_{B,C}=c$.  One also quickly sees that the value of $v_{A,B}$ is inconsequential in establishing the value of $v_{A,C}$:  Eq. \eqref{CST_Velocity} yields $v_{A,C}=c$ regardless of the value of $v_{A,B}$!   Thus, from the perspective of observers in $RFC$, the \textit{Row A} Athenians are observed to be moving in \textit{lock-step} with the \textit{Row B} Athenians, namely all the Athenians in these two rows traveling at light-speed in RFC. This would suggest an answer to the Zeno's question, ``When is $A1$ above $C2$?'':  it is at $t=1\tau$ on clocks in RFC, as shown in Fig. \ref{fig:Fig11}.   

This result may not provide much satisfaction since this perfectly sensible (and expected) solution seems only to buttress the argument that time and space are continuous (again, Eq. \eqref{CST_Velocity} is the CST version of the velocity addition equation)!  Our task then is to show that the rules of special relativity in DST yield a similar result.  If successful, we can finally put to rest debate on this paradox, thereby eliminating one more of the commonly assumed problems with DST that has encumbered the model's acceptance. 

\subsection{Analysis of the Stadium Paradox using a Minkowski Spacetime Diagram}
Before we analyze the resolution of the paradox in DST, let us comment further on the CST case.  Some readers may still have some concerns about the resolution of the paradox in CST that was described in the prior paragraph.  One such concerned can be framed as a question:  

\begin{quote}
\textit{In the proposed solution (Fig. \ref{fig:Fig11}) it is suggested that from $C's$ reference frame rows $A$ and $B$ move in lock-step with each other, but how does one reconcile that with the fact that \textit{Row} $B$ Athenians observe the \textit{Row} $A$ Athenians as moving to their right?  If everyone stops, are the Athenians in rows $A$ and $B$ aligned $A1$ above $B1$, $A2$ above $B2$ ... or are they misaligned?  Mustn't it be one or the other?!}  
\end{quote}

\noindent To answer this question and obtain a deeper understanding of the physics involved that will aid in our later analysis of the situation in DST, let us analyze clock asynchronicity and Minkowski spacetime diagrams. 

\begin{figure}[H]
	\centering\includegraphics[width=11cm]{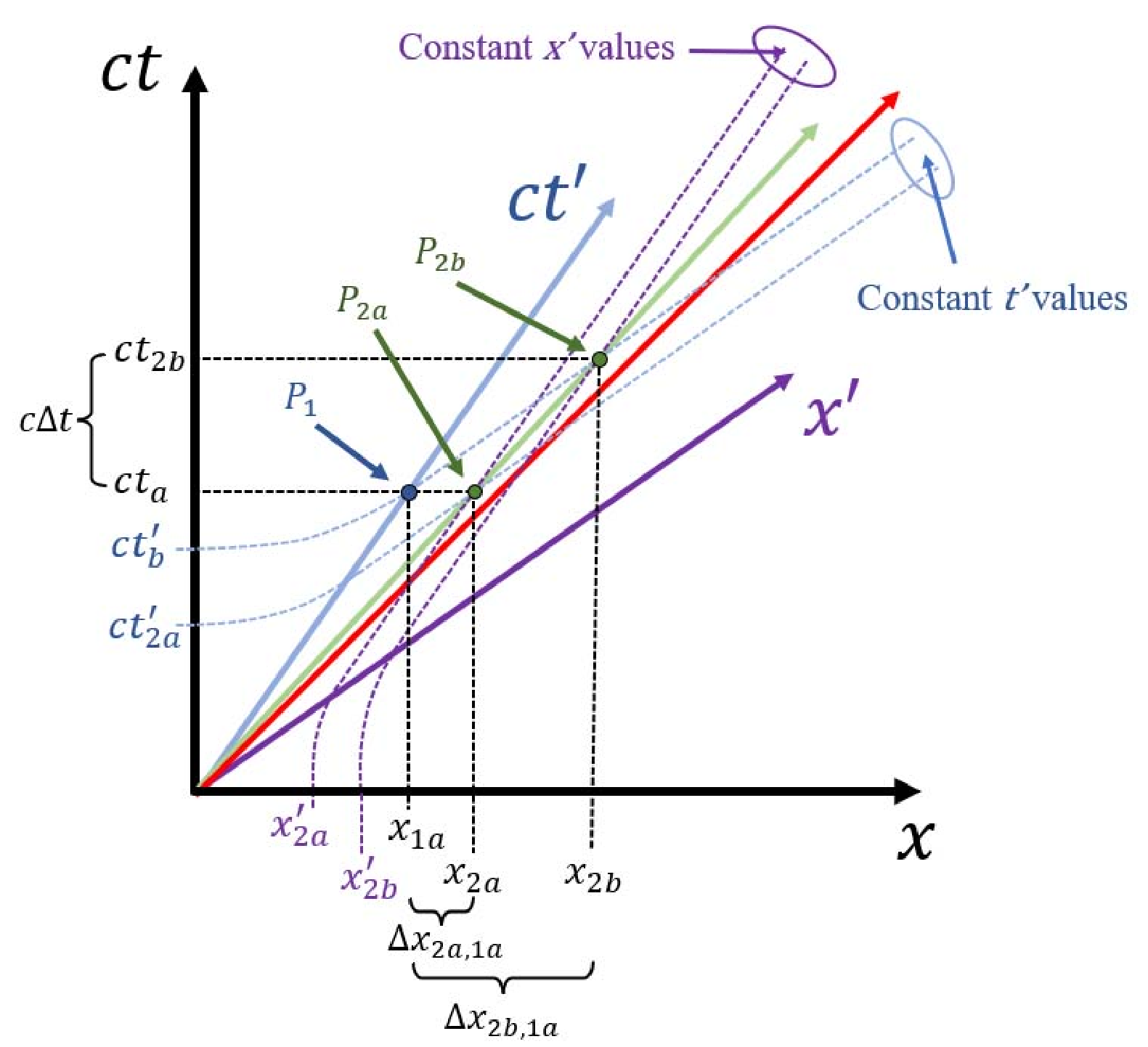}
	\caption{A Minkowski spacetime diagram of two particles:  $P_1$ in blue, and two positions for $P_2$ in green for two different stopping times.  Both particles start at the origins of the two RFs at $t=t'=0$.  RFB has a velocity $v$ relative to RFC and $P2$ has a velocity of $\lambda$ in RFB and $w= \left ( \lambda +v \right ) \left ( 1+v \lambda /c^2 \right )$ in RFC. Computer code to construct these Minkowski spacetime diagrams for arbitrary values of $v$ can be obtained by contacting the author.}
	\label{fig:Fig12}
\end{figure}

Consider two particles in CST: one particle at rest at the origin of RFB (denoted as $P_1$), and a particle (denoted as $P_2$) starting at the origin of RFB at $t'=0$ and traveling at a velocity $\lambda$ in RFB:
\begin{equation} \label{Particle_Position}
	x'_2=\lambda t'
\end{equation} 
We use this relation (Eq. \eqref{Particle_Position}) in the equation for $x'_2$ in the Lorentz transformation equations for this situation (with $\gamma = (1-v^2/c^2)^{-1/2}$):  
\begin{subequations}
	\begin{align}
		x_2 &= \gamma (x'_2 + v t') = \gamma (\lambda t' + vt') = \gamma \left ( \lambda + v \right )t' \label{particle_pos} \\
		t_2 &= \gamma \left ( t' + \frac{v}{c^2}x'_2 \right ) =  \gamma \left (t' + \frac{v}{c^2}\lambda t' \right ) = \gamma \left (1 + \frac{v}{c^2}\lambda \right ) t' \label{particle_time} \\
		x'_2 &= \gamma \left (x_2 - vt \right ) \label{particle_pos_pri} \\
		t'_2 &= \gamma \left (t - \frac{v}{c^2}x_2 \right  ) \label{particle_time_pri}
	\end{align}
\end{subequations}
Of the four equations above, the equation that is important for an analysis of the Stadium Paradox is Eq. \eqref{particle_time}...and of the terms in this equation, it is the clock asynchronicity term that is the most important:
\begin{equation}
	 \Delta t'(v,x') =  \frac{v}{c^2}x'_2 =  \frac{v}{c^2}\lambda t' \label{clock_asymd}
\end{equation}
\noindent As it shown next, it is this term (Eq. \eqref{clock_asymd}) that answers the question of whether the Athenians in rows $A$ and $B$ are aligned or misaligned once the two particles stop moving.   Consider the Minkowski spacetime diagram of the situation shown in Fig. \ref{fig:Fig12}.  The figure is quite complex, so let us provide a detailed list of definitions and an overall key to the figure, given below the figure.

\begin{tcolorbox}[%
    enhanced, 
    breakable,
    colback=blue!3!white,
    colframe=blue!60!white,
     title={Fig. \ref{fig:Fig12} Definitions and Key},
    overlay broken = {
        \draw[line width=.5mm, blue!60!white, rounded corners]
        (frame.north west) rectangle (frame.south east);},
   ]
    RFB has a velocity of $v$ in the $+\hat{x}$ direction relative to RFC.\\
Particle 2 has a velocity of $\lambda$ in RFB, and $w= \left ( \lambda +v \right ) \left ( 1+v \lambda /c^2 \right )$ in RFC.
\vspace{-0.6cm}

\line(1,0){11cm}

\vspace{0.0cm}
\begin{description}
 [font=\normalfont,leftmargin=1.25in,style=multiline]
\setlength\itemsep{0em}
	\item[Green Solid Line:] The world line of $P_2$:  the trajectory of $P_2$ with a slope $w/c$.   The $ct'$ axis is the world line of $P_1$.
	\item[Red Line:] The light line:  the trajectory that a photon has from the perspective of observers in both RFB and RFC.
	\item[Purple Dotted Lines:] Lines of constant $x'$ values and slope of $c/v=1/\beta$ where $\beta = v/c$
	\item[Blue Dotted Lines:] Lines of constant $t'$ values with a slope of $\beta$. 
\end{description}
\vspace{-0.2cm}
\line(1,0){11cm}
\begin{description}
    [font=\normalfont,leftmargin=.4in,style=multiline]
\setlength\itemsep{0em}
   	 \item[$P_1$:] 
		Denotes \textit{Particle 1} that is always at the origin of RFB.  It therefore has a velocity of $v$ in RFC, and a velocity of zero in RFB. 
		\item[$P_{2a}$:] Denotes the position of \textit{Particle 2} on the spacetime diagram when RFC stops $P1$ and $P2$ synchronously.
		\item[$P_{2b}$:] Denotes the position of \textit{Particle 2} on the spacetime diagram when RFB stops $P1$ and $P2$ synchronously.
\end{description}
 \line(1,0){11cm}
\begin{description}
    [font=\normalfont,leftmargin=.4in,style=multiline]
\setlength\itemsep{0em}
	\item[$t_a$:] 
		The time (in RFC coordinates) at which RFC synchronously stopped both $P1$ and $P2$.
	\item[$t'_{b} $:]  The time (in RFB coordinates) at which RFB synchronously stopped both $P1$ and $P2$.
\vspace{0.0cm}
\begin{equation} \label{tprimeb}
t'_{b} = \gamma^{-1} t_a
\end{equation}\vspace{-0.5cm}
\end{description}\vspace{-0.5cm}
 \line(1,0){11cm}
 \begin{description}
    [font=\normalfont,leftmargin=.4in,style=multiline]
	\item[$x_{1a}$:] Where (in RFC coordinates) RFC stopped $P_1$ when synchronously stopping $P1$ and $P2$: 
	\begin{equation}
		x_{1a} = vt_a  \label{x1a}
	\end{equation}
	\item[$x_{2a}$:] Where (in RFC coordinates) RFC stopped $P_2$ when synchronously stopping $P1$ and $P2$: 
	\begin{equation}
		x_{2a} = wt_a=\left ( \frac{\lambda+v}{1+v\lambda/c^2 } \right ) t_a  \label{x2a}
	\end{equation}
	\item[$x_{2b}$:] Where (in RFC coordinates) RFB stopped $P_2$ such that the stopping of $P_1$ and 		$P_2$ is synchronous in RFB.  
	\begin{equation}\label{x2b}
		x_{2b} = wt_{2b} =\left ( v + \lambda \right )t_a 
	\end{equation}
	\item[$t_{2b}$:] When (in RFC coordinates) RFB observers stopped $P2$ such that the stopping of $P_1$ and $P_2$ is synchronous in RFB:
	\begin{equation}
		t_{2b} = \gamma \left ( t'_b + \frac{v}{c^2}x'_{2b} \right ) =\left ( 1 + \frac{v\lambda}{c^2} \right ) t_a 				\label{t2b}
	\end{equation}
\end{description}\vspace{-.25cm}
 \line(1,0){11cm}
 \begin{description}
    [font=\normalfont,leftmargin=.4in,style=multiline]
    	\item[$x'_{1a}$:] Where (in RFB coordinates) RFC stopped $P_1$ when synchronously stopping $P1$ and $P2$: 
	\begin{equation}
		x'_{1a} = 0 \label{x1aprime}
	\end{equation}
	\item[$x'_{2a}$:] Where (in RFB coordinates) RFC stopped $P_2$ when synchronously stopping $P1$ and $P2$:
	\begin{equation}\label{x2aprime}
		x'_{2a} = \lambda t'_{2a} =  \frac{\gamma^{-1} \lambda}{1+\frac{v\lambda}{c^2}} t_a
	\end{equation}
	\item[$x'_{2b}$:] Where (in RFB coordinates) RFB stopped $P_2$ such that the stopping of $P_1$ 			and $P_2$ is synchronous in RFB.
	\begin{equation}\label{x2bprime}
		x'_{2b}=\lambda t'_b = \gamma^{-1} \lambda t_a
	\end{equation}
	\item[$t'_{2a}$] When (in RFB coordinates) RFC stopped $P_2$ such that the stopping of $P_1$ and $P_2$ is synchronous in RFC.
	\begin{equation}
		t'_{2a} = \gamma \left ( t_a - \frac{v}{c^2}x_{2a} \right ) = \frac{\gamma^{-1}}{1+\frac{v\lambda}{c^2}} t_a 					\label{t2aprime}
	\end{equation}
\end{description}\vspace{-0.25cm}
 \line(1,0){11cm}
\begin{description}
    [font=\normalfont,leftmargin=.6in,style=multiline]
	\item[$\Delta x_{2b,1a}$:]  The distance in RFC that $P_2$ would travel \textit{beyond} $P_1$ if the stopping of $P_1$ and $P_2$ were synchronous in RFB.
	\begin{equation}
		\Delta x_{2b,1a} = x_{2b}-x_{1a} = \lambda t_a
	\end{equation}
	\item[$\Delta x_{2a,1a}$:]  The distance in RFC that $P_2$ would travel \textit{beyond} $P_1$ if the stopping of $P_1$ and $P_2$ were synchronous in RFC.
	\begin{equation}
		\Delta x_{2a,1a} = x_{2a}-x_{1a} = \frac{\gamma^{-2}\lambda}{1+\frac{v\lambda}{c^2}}t_a
	\end{equation}
	\item[$\Delta t$:]  The additional time (beyond $t_a$) that observers in RFC should have waited before stopping $P_2$ such that the stopping of $P_1$ and $P_2$ is synchronous in RFB.
	\begin{equation} \label{Delta_t}
		\Delta t = t_{2b}-t_a = \frac{v\lambda}{c^2}t_a
	\end{equation}
\end{description}\vspace{-0.4cm}
\end{tcolorbox}

\vspace{0.4cm}
Equations \eqref{tprimeb}-\eqref{Delta_t} provide us with a comprehensive toolset with which to demonstrate the CST resolution of Stadium.

\vspace{0.5cm}

\noindent \underline{RFC Synchronous Stopping at $t_a$}

There are many ways to use Eqs. \eqref{tprimeb}-\eqref{Delta_t} (with $v,\lambda \rightarrow c$) to demonstrate the CST resolution of Stadium.  The simplest way is to show that when RFC does the motion-stopping we get the following positions and times:
\begin{subequations}
\begin{align}
x_{1a}&=x_{2a}=ct_a \label{Result1} \\
x'_{1a}&=x'_{2a}=0 \label{Result2} \\
t'_{2a}&=0 \label{Result3}
\end{align}
\end{subequations}
\noindent It is easily seen that the positions given by Eq. \eqref{Result1} are obtained from Eqs. \eqref{x1a} and \eqref{x2a} when $v=\lambda = c$;  the positions given by Eq. \eqref{Result2} are obtained from Eqs. \eqref{x1aprime} and \eqref{x2aprime}  with $\gamma^{-1}$ being zero for $v=c$; and the elapsed time of motion for $P_2$ (in RFB time) given by Eq. \eqref{Result3} is obtained from Eq. \eqref{t2aprime} because, again, $\gamma^{-1} = 0$ for $v = c$.

\vspace{0.5cm}

\noindent \underline{RF2 Synchronous Stopping at $t'_{b}$}

We now must show that when RFB does the motion-stopping, we get the following positions and times:
\begin{subequations}
\begin{align}
x_{1a}&=ct_a  \label{Result4} \\
x_{2b}&=2ct_a \label{Result5} \\
t_{2b}&=2t_{a} \label{Result6} \\
x'_{1a}&=0  \label{Result7} \\
x'_{2b}&=c t'_b \label{Result8} 
\end{align}
\end{subequations}
\noindent It is seen that the position given by Eq. \eqref{Result4} is obtained from Eq. \eqref{x1a} as $v\rightarrow c$; the position given by Eq. \eqref{Result5} is obtained from Eq. \eqref{x2b} as $v\rightarrow c$; the elapsed time of motion for $P_2$ in RFC given by Eq. \eqref{Result6} is obtained from Eq. \eqref{t2b}; the position of $P_1$ in RFB given by Eq. \eqref{Result7} is true for all time; and the position of $P_2$ in RFB is obtained from Eq. \eqref{x2bprime} as $\lambda \rightarrow c$.

We see that the special theory of relativity in CST yields the expected results...hence the long-held belief that this paradox was one important nail of many in the coffin for DST.  However, there are a couple of problems with the resolution.  One problem involves associating stopping-times in RFC and RFB.  If motion is stopped for $P1$ and $P2$ by RFB at $t'_b$, then we see from Eqs. \eqref{Result4} and \eqref{Result5} that motion is stopped in RFC at $t_a$ for $P1$ and $2t_a$ for $P2$...but what is the value of $t_a$?  Equation \eqref{tprimeb} states that $t_a = \gamma t'_b=\infty$!  This can be interpreted as making the time (and space) of RFC \textit{inaccessible} to RFB.  In other words, {Row C} Athenians will wait forever for \textit{Row A} and \textit{Row B} Athenians to stop motion, only to be forever dissappointed!  This is not the only problem with the construction of the paradox in CST.  Explained in another work by the author is problem encountered when motion is started:  there is an infinite spreading out of the \textit{Row A} and \textit{Row B} Athenians upon commencement of their motion (i.e., the Athenians in these rows are sequentially whisked away to infinity).  In the next section, we show that the DST version of the resolution yields the same results as the CST version but without $\gamma$ being infinity...thereby making it far superior than the CST resolution. 

\subsection{\label{Stadium_Paradox_DST} The Resolution of the Stadium Paradox in DST}
\begin{figure}[H]
	\centering\includegraphics[width=0.85\textwidth]{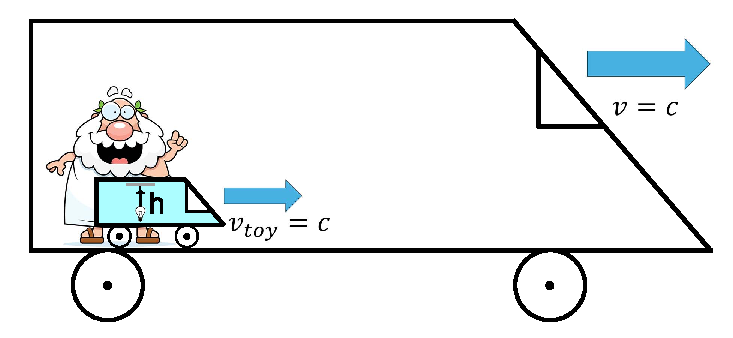}
	\caption{The clever Diogenes believes that he can fool nature.  To get around nature's speed limit of $c$, he simply puts the clocks in his toy train and gives the toy a push.  He believes that observers alongside the tracks will observe the clock to be moving faster than $c$. }  
	\label{fig:Fig13}
\end{figure}

With the relativistic tools appropriate for DST developed in the prior sections of the paper, we are ready to tackle the Stadium Paradox in DST.  From the CST analysis of the paradox, we see that the error in the standard construction of the paradox (Fig. \ref{fig:Fig10}) is the incorrect addition of particle velocities specified in different reference frames, namely velocities added linearly instead of relativistically using Eq. \eqref{CST_Velocity}. To make this error even clearer, consider the equivalent setup shown in Fig. \ref{fig:Fig13} where an Athenian (let us call him Diogenes) attempts to circumvent the laws of nature (specifically, nature's speed limit of $c$) by simply putting the clocks he had in Fig. \ref{fig:Fig3} into a toy train and then giving the train a push.  What we are asked to believe is that this is all it takes to fool nature!  

We have found from the prior sections that there is no prohibition on Diogenes from pushing the toy train towards the front of the cabin at any speed he desires, including $c$...it is just that there will be a disagreement between Diogenes and an Athenian who is stationary beside the tracks (let us call him Zeno) concerning the time at which the position of the toy train is later measured.  This is seen from the DST versions of the normalized LTEs (Eqs. \eqref{Lorentz_Final_DST}, \eqref{GNAST} and \eqref{gamma_expression_DST_a}) evaluated for light-speed motion, namely with $v_{avg}=c$ and given below.  Note that Diogenes is stationary in the \textit{moving frame} (i.e., primed coordinates) and Zeno is stationary in the \textit{rest frame} (i.e., unprimed coordinates). 
\begin{subequations} \label{DiogenesLTEs}
	\begin{alignat}{2}
		\gamma^{-1}_{m} m &= m' + \Delta(c,n') &&= m' + n' \label{Lorentz_Time_H}\\
		\gamma^{-1}_{n} n &= n' + \Delta(c,m') &&= n' + m'  \label{Lorentz_Time_F} \\
		\gamma^{-1}_{m'} m' &= m + \Delta(c,n) &&= m - n \label{Lorentz_Time_I}\\
		\gamma^{-1}_{n'} n' &= n + \Delta(c,m) &&= n - m \label{Lorentz_Time_G}	
	\end{alignat}
\end{subequations}
\noindent where $\Delta(c,j)=\pm j$ with $+$ for Diogenes's reference frame coordinates and $-$ for Zeno's reference frame coordinates.  Also, for light-speed motion, Eq. \eqref{gamma_expression_DST_a}  for $\gamma^{-1}_j$ becomes:
\begin{equation}
	\gamma^{-1}_j j = \Biggl \lceil   \sqrt{2j+1} -1  \Biggr\rceil   \label{lightspeedgammab}
\end{equation}
\noindent where $j$ can be any be any quantity that we need to contract in the calculations below...we will see we need to do so for the following quantities:  $m_{dio}$, $m_{toy}$, $m'_{zeno}$, $m'_{toy}$, $n_{dio}$, $n'_{zeno}$, $n_{toy}$, and $n'_{toy}$.  Equation \eqref{lightspeedgammab} provides important light-speed contractions that we will need later in our analysis of Stadium:
\begin{subequations}
	\begin{align}
	\gamma^{-1}_0 0 &= 0 \label{ContractA}\\
	\gamma^{-1}_1 1 &= 1 \label{ContractB}\\
	\gamma^{-1}_2 2 &= 2 \label{ContractC}
	\end{align}
\end{subequations}

The toy train, Diogenes, and Zeno are aligned at $t=t'=0$, namely $x_{zeno}=x_{dio}=x_{toy}=x'_{zeno}=x'_{dio}=x'_{toy}=0$.  As per the construction of the paradox, the toy train is moving at light-speed in Diogenes's reference frame $x'_{toy}=n'_{dio}\chi$ (in normalized form: $m'_{toy}=n'_{dio}$).  Also, we see from Eq. \eqref{DSTVelocityAddc} that the velocity of the train in RFC (i.e., $v_{toy,RFC}$) is equal to $c$...thus, $x_{toy}=n_{zeno}\chi$ (in normalized form: $m_{toy}=n_{zeno}$).  

Before we look at the motion-stopping from the two perspectives of Diogenes and Zeno, let us look at the two important clock asynchronicity terms in Eqs. \eqref{Lorentz_Time_F} and \eqref{Lorentz_Time_G}, in their unnormalized form, they are:   
\begin{subequations}
	\begin{alignat}{2}
		 \Delta t'_{dio}(v=c,m'_{toy}=n'_{dio}) &= n'_{dio}\tau &&= t'_{dio} \label{deltatdio}\\
		 \Delta t_{zeno}(v=c,m_{toy}=n_{zeno}) &= - n_{zeno}\tau &&= -t_{zeno}  \label{deltatzeno}
	\end{alignat}
\end{subequations}
\noindent These two CATs really tell us the important part of the story:  the ever-present disagreement of stopping-times amongst reference frames.  Zeno will tell Diogenes that he stopped his toy train at twice the time (in Diogenes's reference frame coordinates) he gave himself to move (see Eq. \eqref{deltatdio})...so, of course the toy moved further down the cabin of the train!  Whereas Diogenes will tell Zeno that he measured the position of his toy train much too early---right at the instant Diogenes was going to push it, giving it zero time to travel down the cabin (see Eq. \eqref{deltatzeno})!    No matter when Diogenes or Zeno perform their motion-stopping action, they will err in this way.  However, we see that it is not the fault of Zeno or Diogenes, but of nature's...for she is not so easily tricked.  Let us look into greater detail by assuming the different perspectives (Diogenes's and Zeno's) and calculating stopping times and positions.

\vspace{0.5cm}
\noindent \underline{Diogenes's Motion-Stopping at $n'_{dio}=1$}

Diogenes stops all motion at his normalized time of $n'_{dio}=1$ and he, as expected, observes that his toy train has progressed one hodon to the right ($m'_{toy}=1$) and Zeno has progressed one hodon to the left ($m'_{zeno}=-1$).  

Switching perspectives to Zeno, Eqs. \eqref{Lorentz_Time_H}-\eqref{Lorentz_Time_F} give us the stopping times and positions of Diogenes and his toy train in Zeno's RFC coordinates, namely $m_{dio}$, $n_{dio}$, $m_{toy}$, and $n_{toy}$ when $n'_{dio}=1$:
\begin{subequations} \label{DiogenesLTEs}
	\begin{align}
		\gamma^{-1}_{m_{dio}} m_{dio} &= n'_{dio} = 1  \label{Lorentz_Time_M}\\
		\gamma^{-1}_{n_{dio}} n_{dio} &= n'_{dio} = 1 \label{Lorentz_Time_N} \\
		\gamma^{-1}_{m_{toy}} m_{toy} &= n'_{dio} + n'_{dio} = 2n'_{dio} =2\label{Lorentz_Time_O}\\
		\gamma^{-1}_{n_{toy}} n_{toy} &= n'_{dio} + n'_{dio} = 2n'_{dio} =2 \label{Lorentz_Time_P}
	\end{align}
\end{subequations}
\noindent We use the $\gamma^{-1}$ operation given by Eq. \eqref{lightspeedgammab} (and the specific operations given by Eqs. \eqref{ContractA}-\eqref{ContractC}) with the results above to get:
\begin{subequations} \label{DiogenesLTEsb}
	\begin{alignat}{2}
		m_{dio} &= 1  \label{Lorentz_Time_Q}\\
		n_{dio} &= 1 \label{Lorentz_Time_R} \\
		m_{toy} &= 2 \label{Lorentz_Time_S}\\
		n_{toy} &= 2 \label{Lorentz_Time_T}
	\end{alignat}
\end{subequations}

We see what is troubling Zeno:  from Zeno's perspective, Diogenes performed the measurements of himself and his toy \textit{asynchronously}, having done so on himself at one time (namely, $n=1$ or the unnormalized time of $t=\tau$) and for the toy at $n=2$  (or the unnormalized time of $t=2\tau$).

\vspace{0.5cm}
\noindent \underline{Zeno's Motion-Stopping at $n_{zeno}=1$}

Zeno stops all motion at his normalized time of $n_{zeno}=1$ and he, as expected, observes that both Diogenes and the toy train have progressed one hodon to the right ($m_{dio}=m_{toy}=1$).  

Switching perspectives to Diogenes, Eqs. \eqref{Lorentz_Time_H}-\eqref{Lorentz_Time_F} give us the stopping times and positions of Zeno and the toy train in Diogenes's RFB coordinates, namely $m'_{zeno}$, $n'_{zeno}$, $m'_{toy}$, and $n'_{toy}$ when $n_{zeno}=1$:
\begin{subequations} \label{ZenoLTEs}
	\begin{align}
		\gamma^{-1}_{m'_{zeno}} m'_{zeno} &= -n_{zeno} = -1  \label{Lorentz_Time_Q}\\
		\gamma^{-1}_{n'_{zeno}} n'_{zeno} &= n_{zeno}  = 1 \label{Lorentz_Time_R} \\
		\gamma^{-1}_{m'_{toy}} m'_{toy} &= n_{zeno} - n_{zeno} = 0 \label{Lorentz_Time_S}\\
		\gamma^{-1}_{n'_{toy}} n'_{toy} &= n_{zeno} - n_{zeno} = 0 \label{Lorentz_Time_T}
	\end{align}
\end{subequations}
\noindent We can use the $\gamma^{-1}$ operation given by Eq. \eqref{lightspeedgammab} (and the specific operations given by Eqs. \eqref{ContractA}-\eqref{ContractC}) with the results above to get:
\begin{subequations} \label{DiogenesLTEsb}
	\begin{alignat}{2}
		m'_{zeno} &= -1  \label{Lorentz_Time_Q}\\
		n'_{zeno} &= 1 \label{Lorentz_Time_R} \\
		m'_{toy} &= 0 \label{Lorentz_Time_S}\\
		n'_{toy} &= 0 \label{Lorentz_Time_T}
	\end{alignat}
\end{subequations}

We now see what is troubling Diogenes:  from Diogenes's perspective, Zeno has performed the motion-stopping of himself (Zeno) correctly, but he has stopped the toy before any time has passed in Diogenes's reference frame, giving no time (namely, $n'_{toy}=0$) for the toy to move down the cabin of the train.

\section{\label{sec:SRTopics} Discussion}
Two issues remain to be discussed.  The first is the interesting phenomenon of the emergence of positional and temporal uncertainties in what was once fully deterministic physics (i.e., conventional special relativity in CST).  The second involves particle-particle interactions in DST.

\subsection{\label{sec:Position_Uncertainty}Positional and Time Uncertainty}
An interesting consequence of DST is the positional and temporal uncertainties in the stopping points of otherwise identically prepared systems (i.e., particles).  Giving an example is the best way to describe this effect.   

Consider a particle traveling at $v_{avg}=0.99c$.  Let the particle start traveling at $t'=0$ along a path of length $m = 50$ (relative to $\chi$); $m=50$ is the rest frame length, the corresponding contracted length is $m'=10$ (see Table \ref{table:Table6}).  If the particle travels uninterrupted, then it will travel the rest-frame distance of $m=50$ distance in a time of $n=51$ (relative to $\tau$); the corresponding contracted time is $n'=11$.  Thus, the moving particle in its reference frame will experience a greatly contracted length of $m'=10$ in a time of $n'=11$.

Now consider what happens when the particle stops in its reference frame at $n'=11$.  We see from Table \ref{table:Table6} that there will be a range of spatial positions and times throughout which the particle can occupy.   This is because the moving particle can stop only on one of the time ticks \textit{in its reference frame}---this is the finest resolution that is possible in DST!   However, when the particle's clock ticks $n'=11$, this single tick of its clock spans a range of ticks in the at-rest reference frame of $n=51 \rightarrow 61$.  The corresponding length in the at-rest frame is $m=50 \rightarrow 60$; the moving particle will have absolutely no control of when in this time duration and where in this spatial distance it inhabits when it stops moving at $m'=10$ and $n'=11$.  We could set up an experiment with $N$ identically prepared systems, each with the same velocity, and start and stop their motion identically in its (RF2's) reference frame.  However, even if this procedure is done perfectly, we will find that the particles populate a range of positions in the at-rest reference frame.  This result is a natural consequence of the combination of discretization of spacetime and the relativistic contractions of space and time.
\begin{table} [H]
\caption{The time dilation and length contraction associations for two reference frames with a relative velocity of $v_{avg}=0.99c$.}\label{table:Table6}
\begin{minipage}[t]{0.32\textwidth}\vspace{0pt}
\begin{tabular}{c|c|c|c}
\toprule
$n$ & $n'$ & $m$ & $m'$ \\
\midrule
1 & 1 & 0 & 0   \\
2 & 2 & 1 & 1   \\
3 & 3 & 2 & 2   \\
4 & 3 & 3 & 2   \\
5 & 3 & 4 & 2   \\
6 & 4 & 5 & 3   \\
7 & 4 & 6 & 3   \\
8 & 4 & 7 & 3   \\
9 & 5 & 8 & 4   \\
10 & 5 & 9 & 4   \\
11 & 5 & 10 & 4   \\
12 & 5 & 11 &  4  \\
13 & 5 & 12 & 4   \\
14 & 6 & 13 & 5   \\
15 & 6 & 14 & 5   \\
16 & 6 & 15 & 5   \\
17 & 6 & 16 & 5   \\
18 & 6 & 17 & 5   \\
19 & 7 & 18 &  6  \\
20 & 7 & 19 &  6  \\
21 & 7 & 20 &  6  \\
22 & 7 & 21 &  6  \\
23 & 7 & 22 &  6  \\
24 & 7 & 23 &  6  \\
25 & 7 & 24 &  6  \\
\vdots & \vdots & \vdots & \vdots \\ 
\bottomrule
\end{tabular}
\end{minipage} \hfill
\begin{minipage}[t]{0.32\textwidth}\vspace{0pt}
\begin{tabular}{c|c|c|c}
\toprule
$n$ & $n'$ & $m$ & $m'$ \\
\midrule
\vdots & \vdots & \vdots & \vdots  \\
26 & 8  & 25  & 7   \\ 
27 & 8  & 26  & 7   \\ 
28 & 8  & 27  & 7   \\ 
29 & 8  & 28  & 7   \\ 
30 & 8  & 29  & 7   \\ 
31 & 8  & 30  & 7   \\ 
32 & 8  & 31  & 7   \\ 
33 & 9  & 32  & 8   \\ 
34 & 9  & 33  & 8   \\ 
35 & 9  & 34  & 8   \\ 
36 & 9  & 35  & 8   \\ 
37 & 9  & 36  & 8   \\ 
38 & 9  & 37  & 8   \\ 
39 & 9  & 38  & 8   \\ 
40 & 9  & 39  & 8   \\ 
41 & 9  & 40  & 8   \\ 
42 & 10  & 41  & 9   \\ 
43 & 10  & 42  & 9   \\ 
44 & 10  & 43  & 9   \\ 
45 & 10  & 44  & 9   \\ 
46 & 10  & 45  & 9   \\ 
47 & 10  & 46  & 9   \\ 
48 & 10  & 47  & 9   \\ 
49 & 10  & 48  & 9   \\ 
\vdots & \vdots & \vdots & \vdots \\ 
\bottomrule
\end{tabular}
\end{minipage}
\begin{minipage}[t]{0.32\textwidth}\vspace{0pt}
\begin{tabular}{c|c|c|c}
\toprule
$n$ & $n'$ & $m$ & $m'$ \\
\midrule
\vdots & \vdots & \vdots & \vdots  \\
50 & 10  & 49  & 9   \\ 
51 & 11 & 50 & 10   \\ 
52 & 11 & 51 &  10  \\ 
53 & 11 & 52 &  10  \\ 
54 & 11 & 53 & 10   \\ 
55 & 11 & 54 & 10   \\ 
56 & 11 & 55 & 10   \\ 
57 & 11 & 56 & 10   \\ 
58 & 11 & 57 & 10   \\ 
59 & 11 & 58 & 10   \\ 
60 & 11 & 59 & 10   \\ 
61 & 11 & 60 & 10   \\ 
62 & 12 & 61 & 11   \\
63 & 12 & 62 & 11   \\
64 & 12 & 63 & 11   \\
65 & 12 & 64 & 11   \\
\vdots & \vdots & \vdots & \vdots  \\
\bottomrule
\end{tabular}
\end{minipage}
\end{table}

\subsection{\label{sec:Particle Interaction}Particle Interaction}
The two parts of our work on DST, \textit{Part 1} and this paper (\textit{Part 2}), constitute only the start of a complete development of a DST model.  Many other sections could be added to this paper that address additional topics, such as Zeno's other paradoxes, gravity within DST, and additional discussions on motion and time in DST.  To keep this paper of manageable length, however, we will address these topics in other papers and end the discussion with a common question about particle interaction in DST.

\begin{figure}[H]
	\centering\includegraphics[width=0.55\textwidth]{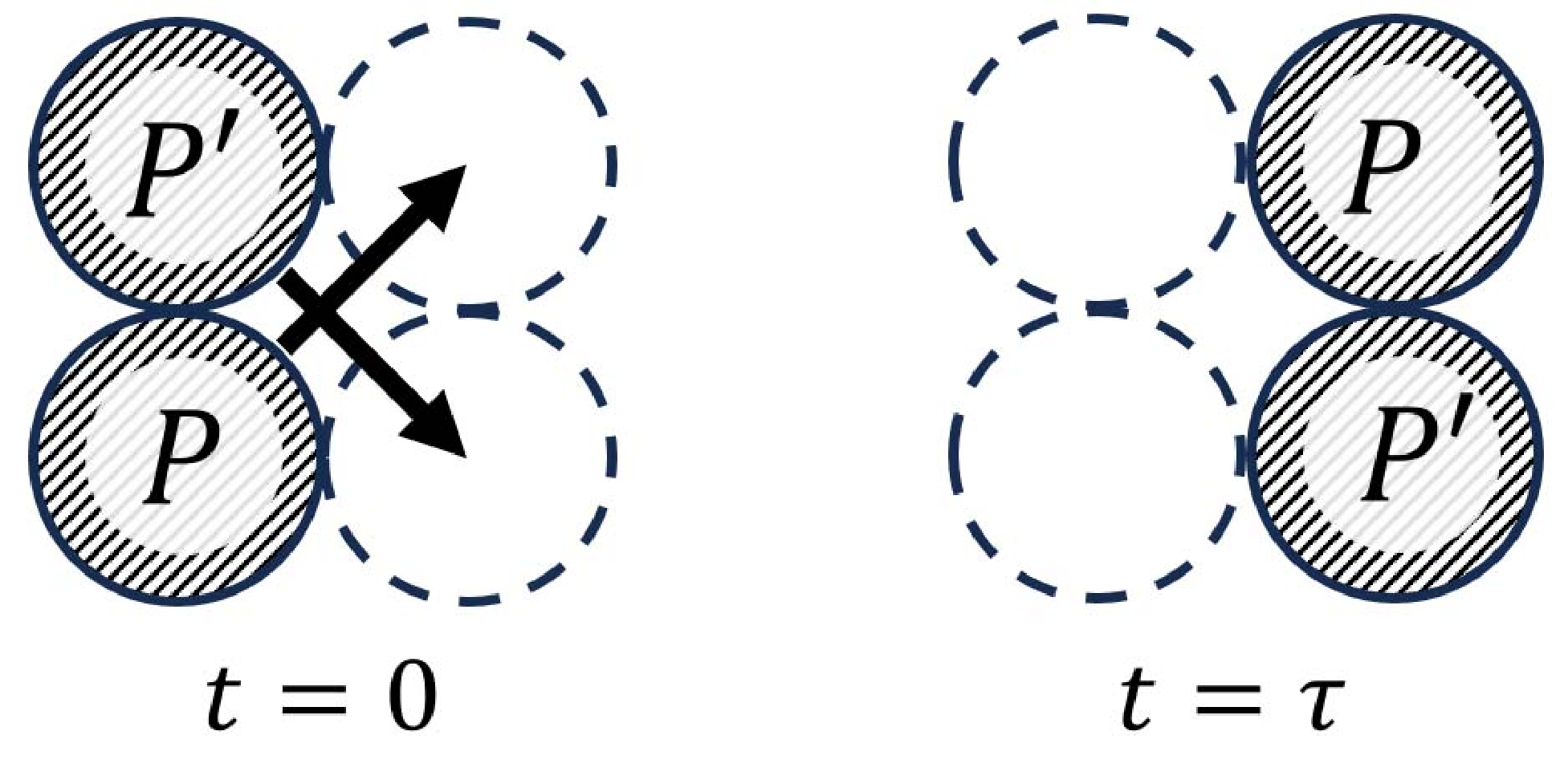}
	\caption{A problematic construction of particle travel in lattice-based DST models.  The imposition of a grid creates certain scenarios of particle motion where particles can cross paths but not interact.   \textbf{Left:} At $t=0$ two particles are vertically adjacent and traveling at $\pm 45^\circ$ directions.  \textbf{Right:} At $t=\tau$, they will have crossed paths but never having overlapped...therefore never interacting. There is little to no chance of this happening in the grid-less \textit{Isotropic Model of DST} discussed in this work because such perfect alignment (at $t=0$ or $t=\tau$) is highly unlikely. }
	\label{fig:Fig14}
\end{figure}

A question that has been asked of us is how the \textit{Isotropic Model}, or any DST model, allows for particle-particle interaction.  The problem-statement is constructed in a way shown in Fig. \ref{fig:Fig14} where a grid is first drawn and two traveling particles are shown.  One particle has a starting position of $(x,y)=(0,0)$ and traveling at a $45^\circ$ at a speed of $c$.  The other particle has a starting position of $(x,y)=(0,\chi)$ and is traveling at a $-45^\circ$ at a speed of $c$ (with $\chi$ being the hodon).  The configuration of the particles at $t=0$ is shown on the left side of Fig. \ref{fig:Fig14}; the configuration of the particles at $t=\tau$ (with $\tau$ being the chronon) is shown on the right side of Fig. \ref{fig:Fig14}.  The statement that is then made is:  \textit{the particles have never overlapped and therefore will not interact}.   As constructed, the statement is true as it pertains to the particle configuration shown in Fig. \ref{fig:Fig14}.  Let us point out the error of the problem-statement and construction:  \textit{the imposition of the grid.}  The very foundational concept of the \textit{Isotropic Model of DST} was the rejection of any such grid!  In the \textit{Isotropic Model} there is not a regular array of spatial states that can be occupied by particles.  Numerous problems, from anisotropy to particle-particle interaction, are caused by the erroneous \textit{a priori} imposition of a grid.  And while the absence of a grid in the \textit{Isotropic Model} may make it difficult for mathematicians and scientists to use/extend the model, such efforts promise to bear fruit.

\section{\label{sec:Conclusion}Conclusion}
In this paper, \textit{Part 2} of our series on discrete spacetime, we have completed the development of the \textit{Isotropic Model} of discrete spacetime.  First, we reviewed \textit{Part 1} where we developed a new formula for calculating distances and then derived the correct equations describing length contraction and time dilation in DST.  We next developed the Lorentz transformation equations appropriate for DST.  The time transformation equation required a careful analysis of clock asynchronicity in DST.  We then derived the relativistic velocity addition equation for DST.  We then gave examples of Minkowski spacetime diagrams.  We finished by showing how the new relativistic tools for DST allowed for a resolution of Zeno's Stadium Paradox.  With this done, we have satisfactorily addressed four of the five important problems historically associated with DST:  adherence to relativity, causality, distance calculations, and the Zeno's Stadium Paradox.  One issue remains that must be studied further:  conservation of energy and momentum.    A careful analysis is needed to determine if, as hypothesized by the author and others, energy and momentum are emergent properties and occur due to the collective interactions of between particles.  If this can be shown, then the last obstacle for the acceptance of DST will have been cleared away.
%
%
%
%
%
%
%
%



\begin{flushright}
	DAVID CROUSE\\
	Professor \\ Department of Electrical\\ and Computer Engineering\\
	Clarkson University\\
	United States of America\\
	\href{mailto:dcrouse@clarkson.edu}{dcrouse@clarkson.edu}\\
	\url{www.clarkson.edu/people/david-crouse}
	
\end{flushright}

\end{document}